\documentclass[12pt,a4paper]{article}

\usepackage{soul}        

\usepackage{amsmath,amssymb,epsf,epsfig}
\usepackage{array}
\usepackage{fancybox}
\allowdisplaybreaks[1]

\usepackage[usenames]{color}
\usepackage{amsfonts,bm}
\usepackage{graphicx}
\usepackage{enumerate}

\usepackage{parskip}     
\usepackage[
      colorlinks=true,
      linkcolor=blue,
      urlcolor=blue,
      filecolor=blue,
      citecolor=blue,
      pdfstartview=FitH,
      pdftitle={},
      pdfauthor={},
      pdfsubject={},
      pdfkeywords={},
      pdfpagemode=UseNone,
      bookmarksopen=true,
      ]{hyperref}
\usepackage[all]{hypcap}     

\numberwithin{equation}{section}

\newcommand{\be}{\begin{equation}}
\newcommand{\ee}{\end{equation}}
\newcommand{\bea}{\begin{eqnarray}\displaystyle}
\newcommand{\eea}{\end{eqnarray}}

\def\beq{\begin{equation}}
\def\eeq{\end{equation}}
\def\beqa{\begin{eqnarray}}
\def\eeqa{\end{eqnarray}}
\def\bet{\begin{tabular}}
\def\eet{\end{tabular}}
\def\bs{\begin{split}}
\def\es{\end{split}}

\def\a{\alpha}

\def\m{\mu}

\def\n{\nu}

\def\one{{\hbox{\kern+.5mm 1\kern-.8mm l}}}
\def\zero{{\hbox{0\kern-1.5mm 0}}}

\definecolor{orange}{rgb}{1,0.5,0}

\newcommand{\ket}[1]{{\,| {#1} \rangle}}

\newcommand{\bean}{\begin{eqnarray*}}
\newcommand{\eean}{\end{eqnarray*}}

\begin{document}

\hfill \hbox{DFPD-12-TH-15} 

\vspace{-0.5cm}
\hfill \hbox{QMUL-PH-12-16}

\vspace{2cm}

\centerline{
\LARGE{ \textsc{Perturbative superstrata}} }



\vspace{1.5cm}

\centerline{    
  \textsc{ Stefano Giusto$^{1,2}$, ~Rodolfo Russo$^3$}  }

\vspace{0.5cm}

\begin{center}
$^1\,${Dipartimento di Fisica e Astronomia ``Galileo Galilei'',\\
Universit\`a di Padova,\\ Via Marzolo 8, 35131 Padova, Italy\\
}
\end{center}

\vspace{0.1cm}
 
\begin{center}
$^2\,${INFN, Sezione di Padova,\\
Via Marzolo 8, 35131, Padova, Italy}
\end{center}
\vspace{0.1cm}

\begin{center}
$^3\,${Queen Mary University of London,\\
Centre for Research in String Theory, School of Physics and Astronomy\\
Mile End Road, London E1 4NS, UK\\
}
\end{center}

\vspace{0.2cm}

\begin{center}
{\small stefano.giusto@pd.infn.it, ~~r.russo@qmul.ac.uk}
\end{center}

\vspace{1cm}

\centerline{ 
 \textsc{ Abstract}}

\vspace{0.2cm} {\small We study a particular class of D-brane bound
  states in type IIB string theory (dubbed ``superstrata'') that 
  describe microstates of the 5D Strominger-Vafa black hole. By using the
  microscopic description in terms of open strings we probe these
  configurations with {\em generic} light closed string states and
  from there we obtain a linearized solution of six-dimensional supergravity 
  preserving four supersymmetries. We then discuss two generalizations of the
  solution obtained which capture different types of non-linear
  corrections. By using this construction, we can provide the first
  explicit example of a superstratum solution which includes the
  effects of the KK-monopole dipole charge to first order.}
        
\thispagestyle{empty}

\vfill
\eject

\setcounter{page}{1}

\section{Introduction} \label{sec:introduction}

In string theory supersymmetric black holes are realized as bound
states at threshold of many basic constituents, such as perturbative
string states and branes. In~\cite{Strominger:1996sh,Callan:1996dv}
this picture was used to count in concrete examples the degeneracy of
the configurations that have the same (three) conserved charges. In
particular, the setup studied in those papers is type IIB string
theory compactified on a $S^1$ of radius $R\gg \sqrt{\alpha'}$ times a
string-sized four manifold, which is either $T^4$ or $K_3$. In the
large charge limit the microscopic counting matches perfectly the
Bekenstein-Hawking entropy of a black hole solution with the same
charges. However, while the study of the bound state degeneracy is
performed at zero string coupling $g_s$, the gravitational
backreaction of the string/brane bound state, an thus its connection
to black holes, is manifest only when $g_s N$ is big, where $N$
roughly indicates the number of elementary constituents of each type
present in the bound state. The presence of four preserved
supercharges is crucial for connecting the degeneracy of the
configuration at $g_s=0$ with the black hole entropy derived from the
black hole geometry. In~\cite{Lunin:2001fv, Lunin:2001jy,Lunin:2002qf}
a new line of research was initiated with the aim to understand the
gravitational backreaction of the different configurations (known as
microstates), which at the level of the free theory account for the
bound state degeneracy. One of the aims of this programme is to
understand whether it is possible to give a supergravity description
of each microstate in the limit when $g_s$ is small, but $g_s N$ is
finite. Even if this geometric description fails when approaching the
bound state under analysis, the really important question is whether
this happens before a horizon is formed or not.

Even though there has been continuous progress over the last ten years
and many works have been published on the subject (see for instance
the review
articles~\cite{Mathur:2005zp,Bena:2007kg,Skenderis:2008qn,Mathur:2008nj,Balasubramanian:2008da,
  Chowdhury:2010ct, Mathur:2012zp}), a complete understanding of the
3-charge microstate geometries relevant for the black hole
studied in~\cite{Strominger:1996sh,Callan:1996dv} is still lacking. The
question of whether the generic microstate can be described by a horizonless geometry 
is still open. However very far away from the location
of the bound state, the geometry will be certainly described by a 5D
Minkowski metric times the compact space plus tiny corrections. There
are cases where this asymptotically flat part can be glued with a
smooth geometry describing the gravitational backreaction of
a particular 3-charge bound state at any value of the radial distance
$r$, see~\cite{Lunin:2004uu,Giusto:2004id,Giusto:2004ip} for some of
the first examples of such geometries. These configurations appear as
1/8-BPS solutions of the standard type IIB supergravity equations, but
with some striking geometric properties. In this case one can check
explicitly that no horizons develop inside the throat of the geometry
and the sources appear to have dissolved into fluxes of the
supergravity gauge fields, which makes the classical solution regular
even in the interior. A more general class of 1/8-BPS solutions with
the same features was constructed in
~\cite{Bena:2005va,Berglund:2005vb}.  The set of 3-charge microstates
with a known gravitational dual was further extended
in~\cite{Ford:2006yb}: the main novelty of this new class of
backgrounds is that the large $S^1$ present in the compact space plays
a non-trivial role and the solutions are genuinely six-dimensional.

In this paper we will follow the approach
of~\cite{Giusto:2009qq,Black:2010uq,Giusto:2011fy} and use the
microscopic description of the bound state in terms of D-branes and
open string states to derive the geometric properties of 3-charge
microstates in particular limits. The basic idea of this approach is
that the couplings of the bound state with the massless closed string
sector of the theory (i.e. the supergravity fields) are described in
the underlying Worldsheet Conformal Field Theory (WCFT) by a set of
correlation functions with disk topology. The conditions imposed
on the boundary of the disk and the possible presence of open string
states define the particular microstate under analysis. The WCFT
correlators we are interested in will always have a single external
closed string state. These correlators are directly related to the
backreaction of the D-brane bound state for the supergravity field
corresponding to the closed string state considered. Physically this
closed string is the probe used to explore the backreacted
geometry. In~\cite{Giusto:2009qq,Black:2010uq,Giusto:2011fy} this
probe was always taken to have zero momentum in all compact
directions. Of course, the information derived in this way cannot
distinguish between localized or smeared configurations and the
supergravity solutions derived could be interpreted entirely in a
(non-minimal) five-dimensional supergravity~\cite{Giusto:2012gt}.

We wish to revisit this analysis by allowing closed string probes that
are localized in the large $S^1$ of the compact space, or in other
words that have momentum along this $S^1$.  In particular, we
will still focus on the D-brane configurations discussed
in~\cite{Giusto:2011fy}, but we wish to explore the geometry in a
finer detail by using the more general probes mentioned above (of
course the same generalization can be done for other D-brane
configurations). Our analysis shows that, in the D1-D5 frame, the
three charge microstates behave differently from the two charge ones:
while in this second case the use of localized probes does not give
any new information, the three charge microstates seem to always have
a non-trivial dependence on the $S^1$ coordinate. A purely 5D geometry
is obtained only if we focus just on delocalized probes along the
$S^1$ or, in other words, we smear over the $S^1$. 
In principle, this process of smearing can
induce spurious singularities which would be absent in the complete
geometry and it is an interesting open problem to understand exactly
when this happens. The smeared configurations are likely to be described
by exotic or non-geometric configurations, of the type studied in~\cite{deBoer:2010ud,deBoer:2012ma}.

The bound states we are interested in are constructed at the
microscopic level by taking D5 branes wrapped on the whole compact
space and D1 branes wrapped on the $S^1$ and by giving them an
identical profile describing a vibration in the transverse non-compact
space. As usual, in order to preserve some supercharges, the functions
$f^i$ describing the shape of the D-branes in the transverse directions can
depend only on either one of the two light-cone coordinates, $v$ and $u$, 
constructed out of time and the $S^1$ coordinate, but
not on both. 
In order to have a real bound state, one should switch on a
non-trivial KK-monopole dipole charge which at the microscopic level
corresponds to give a non zero vacuum expectation value (vev) to some
D1/D5 open string states~\cite{Giusto:2009qq}.
It was argued in~\cite{Bena:2011uw}, mostly based on supersymmetry arguments,
that this class of bound states should be described by smooth supergravity configurations 
parameterized by arbitrary functions of two variables, that were dubbed ``superstrata''.  
The construction of the exact supergravity solutions for superstrata is an important open problem: the first steps in this direction were taken in~\cite{Bena:2011dd,Niehoff:2012wu} (building on previous supergravity results of~\cite{Gutowski:2003rg,Cariglia:2004kk}), which derived exact supergravity solutions representing a superposition of D1 and D5 branes with generic oscillation profiles but no KK-monopole dipole charge. In both the WCFT and
supergravity approaches it is easier to start by treating the KK-monopole dipole
charge perturbatively. The main goal of this paper is to provide the
first explicit example of a solution which includes the effects of the
KK-monopole dipole charge to first order. It is interesting to
notice that the supergravity solutions emerging from the simplest
D-brane configurations studied here by following~\cite{Giusto:2011fy}
do not fall in the ansatz considered in~\cite{Bena:2011dd}. It should
be possible to engineer a D-brane configuration whose backreaction
contains only the fields of the restricted ansatz~\cite{Bena:2011dd},
but this will involve a more complicated choice of the open string
vev's defining microscopically the bound state. Thus simpler
microscopic configurations correspond to more complicated supergravity
solutions and vice versa. This might be somewhat unexpected or be just
a result of the fine tuning required at the microscopic level to
cancel the extra dipoles which are allowed by supersymmetry but are
absent in the ansatz~\cite{Bena:2011dd}.

The paper is structured as follows. In Section~\ref{sec:ansatz} we
generalize the ansatz discussed in~\cite{Giusto:2012gt} so as to adapt
it to the $v$-dependent case we are interested in. The full list of
constraints imposed by supersymmetry and the equations of motion on
the functions appearing in the ansatz is not known. Work is in
progress to derive these equations from first
principles~\cite{GMPR}. However it is not difficult to start from the
equations derived in~\cite{Bena:2011dd} and generalize them at the
linearized level needed for our analysis. In
Section~\ref{sec:disc-rev} we collect the results for the 1-point
string correlators mentioned above and extract the geometric
information we need by comparing them against the ansatz of
Section~\ref{sec:ansatz}. In Section~\ref{sec:disc-rev} we provide a
first generalization of the results obtained from string theory and
show that it is natural to write the linearized supergravity
configuration in terms of a set of simple scalar functions and
1-forms. In particular the 1-form $\beta$ captures the KK-monopole
dipole charge of the configuration; in the diagrammatic language of
string amplitudes $\beta$ is related to the disk amplitudes with the
insertion of one $g_{ui}$ graviton and an arbitrary number of twisted
open string vertices. We show that the linearized equations in the
bulk are satisfied if we assume some simple properties for the basic
building blocks of the supergravity configuration. In particular they
must enjoy the same harmonic and duality constraints that, in the
perturbative string approach of Section~\ref{sec:disc-rev}, follow
from the BRST invariance of the open string
vertices~\cite{Giusto:2009qq}. In Section~\ref{sec:BGSW} we focus on
the special case where the functions $f^i$ describing the D-brane
profile have periodicity $2\pi R$ (this needs not to be the case when
the D-branes are multiply wrapped as it happens for generic three
charge states). This is the class of microstates studied from a
supergravity point of view in~\cite{Bena:2011dd}. We show that this
case is more easily studied in a coordinate system where the metric
for the non-compact space is conformally flat, but the 10D metric is
not asymptotically Minkowski. In this frame the supergravity equations
take a particularly simple form. This allows us to present a further
generalization and obtain an explicit solution which includes the
non-linear terms in the D1 and D5 charges, but is still linear in the
KK-monopole dipole charge. In Section~\ref{sec:conclusions} we present
our conclusion by discussing some possible further developments and
the connections of our approach with recent supergravity literature on
the subject.

\section{The supergravity ansatz} \label{sec:ansatz}

Eq.~(2.8) of~\cite{Giusto:2012gt} contains an explicit ansatz for type
IIB supergravity compactified on $S^1 \times T^4$ which preserves 4
supercharges provided that the conditions (2.9)--(2.11) of that paper
are satisfied. We now wish to extend that ansatz to the case where all
functions and forms appearing in the various fields can depend on $v$,
besides the $R^4$ coordinates $x^i$, with
\be\label{lc-coord}
v = \frac{t+y}{\sqrt{2}}\,,\quad u = \frac{t-y}{\sqrt{2}}\,,
\ee
where $t$ and $y$ indicate the coordinates along the time and the
$S^1$ direction respectively. We will also rephrase the ansatz by
using the language of~\cite{Gutowski:2003rg,Bena:2011dd}. Then the
string frame metric takes the following form
\be
ds^2 = -2\,\frac{\alpha}{\sqrt{Z_1 Z_2}}\,(dv+\beta)\,\Bigl(du+\omega + \frac{\mathcal{F}}{2}(dv+\beta)\Bigr)+\sqrt{Z_1 Z_2}\,ds^2_4 + \sqrt{\frac{Z_1}{Z_2}}\,ds^2_{T^4}\,,
\label{metrica}
\ee
where $\alpha$, ${\cal F}$ and the $Z_I$'s are functions depending on
$v$ and the $R^4$ coordinates $x^i$, while $\omega$ and $\beta$ are
1-forms on $R^4$ but can depend on $v$ as well. The two 4D metrics
$ds^2_{4}$ and $ds^2_{T^4}$ indicate the non-compact and the $T^4$
metric respectively. For the time being we allow for a general
$v$-dependent $R^4$ metric $h_{ij}$ and for the sake of simplicity
take the torus to be perfectly cubic
\be\label{g4dm}
ds^2_4 = h_{ij}\,dx^i\,dx^j\,,\quad ds^2_{T^4} = (dz^1)^2 + \ldots + (dz^4)^2\,. 
\ee
The ansatz for the remaining type IIB supergravity fields is written
in terms of a function $Z_4$ related to $\alpha$ by
\be
\alpha = \Bigl(1-\frac{Z_4^2}{Z_1 Z_2}\Bigr)^{-1}\,,
\ee
the 1-forms $a_1$ and $a_4$, the 2-forms $\gamma_2$ and $\delta_2$ and the 3-form
$x_3$. All these ingredients, except for $x_3$, already appear in the
$v$-independent ansatz~\cite{Giusto:2012gt}. Thus it is useful to
summarize the redefinition necessary to map the conventions of that
paper with the conventions used here:
\begin{equation}\label{redefa2b}
\beta=\frac{\hat{a}_3}{\sqrt{2}}\,,~~ \omega = \sqrt{2}\,\hat{k} - \frac{\hat{a}_3}{\sqrt{2}}\,,~~ \mathcal{F} = -2\,(\hat{Z}_3-1)\,,~~
a_1 = \sqrt{2}\, \hat{a}_1\,,~~ a_4 = \sqrt{2}\, \hat{a}_4\,,
\end{equation}
where the hatted quantities are those appearing
in~\cite{Giusto:2012gt}. Now we can complete the list of fields
appearing in our ansatz. For the dilaton we take
\be
e^{2\phi} = \alpha\,\frac{Z_1}{Z_2}\,,
\ee
and the NS-NS 2-form is
\be
B^{(2)} = -\frac{\alpha\,Z_4}{Z_1 Z_2}\,(du+\omega)\wedge (dv+\beta) + a_4\wedge (dv+\beta) + \delta_2\,.
\ee
Finally the Ramond-Ramond (RR) forms are
\bea \nonumber
C^{(0)} &=& \frac{Z_4}{Z_1}\,,\\
C^{(2)} &=&  -\frac{\alpha}{Z_1}\,(du+\omega)\wedge (dv+\beta) +
a_1\wedge (dv+\beta) + \gamma_2\,,\label{RRa} \\ \nonumber
C^{(4)}&=& \frac{Z_4}{Z_2}\,dV_{T^4} -\frac{\alpha\,Z_4}{Z_1 Z_2}\,\gamma_2\wedge (du+\omega)\wedge(dv+\beta)+ x_3\wedge (dv+\beta)\,,
\eea
where $dV_{T^4} = dz^1 \wedge dz^2 \wedge dz^3 \wedge dz^4$.

This generalizes the ansatz studied in~\cite{Bena:2011dd} by adding the fields 
$B^{(2)}$, $C^{(0)}$ and $C^{(4)}$, which, in the language of 6D supergravity, should
correspond to the addition of an extra tensor multiplet on top of the gravity and tensor multiplet 
already present in~\cite{Bena:2011dd}. We
leave the detailed analysis of the constraints imposed by
supersymmetry and the equations of motion in this more general set up to a forthcoming
paper~\cite{GMPR}. However, in most of this paper we will be working in the approximation in which only linear terms in the expansion of the geometry around flat space are kept: at this linearized level
 the new fields present in the ansatz above basically
decouple from those already present in the ansatz used
in~\cite{Bena:2011dd}. So we can use the equations discussed in that 
paper and supplement them with a set of linearized equation for
$Z_4$, $a_4$, $\delta_2$ and $x_3$. 

Let us denote by $d$ the differential with respect to the $R^4$ coordinates
and define
\be
 D \equiv d - \beta\wedge \partial_v\,. 
\ee
The first conditions are on the 1-form $\beta$ and the 4D metric
$ds^2_4$: $\beta$ has to satisfy
\be\label{bstarb}
D\beta = \star_4 D\beta\,,
\ee
where the $\star_4$ represents the Hodge star according the the $R^4$
metric $h_{ij}$; the Hodge dual with respect to the flat  $R^4$ will instead be
denoted as $*_4$. The metric $ds^2_4$ has to be ``almost hyperkahler",
which means that there exist three 2-forms $J^{(A)}\equiv
\frac{1}{2}\,J^{(A)}_{ij}\,dx^i\wedge dx^j$, with $A=1,2,3$,
satisfying 
\be\label{eq:Jeq} 
J^{(A)} \wedge J^{(B)} = - 2
\delta^{AB} \star_4 1  ~,~~~ 
d J^{(A)} = \partial_v (\beta\wedge J^{(A)})\,.
\ee
This implies that the $J$'s are anti-self-dual with respect to the
star $\star_4$ defined above: 
\be
J^{(A)} = - \star_4 J^{(A)}\,.
\ee 
As usual, by raising one index and
choosing an appropriate ordering, we can define three almost complex
structures
\begin{equation}
  \label{eq:acs}
  {J^{(A)\, i}}_k\,{J^{(B) \,k}}_j = \epsilon^{ABC}\,{J^{(C)\, i}}_j -
  \delta^{AB}\,{\delta^i}_j~.
\end{equation}
For later use, by using the complex structures we define a new
anti-self-dual 2-form $\psi$
\be\label{psidef}
 \psi \equiv \frac{1}{8} \,\epsilon^{ABC}\,J^{(A)\,ij}\,{\dot{J}^{(B)}}_{ij}\,J^{(C)}\,,
\ee
where the dot indicates the derivative with the respect to $v$. 

Let us
now consider the equations for the part of the ansatz determining the
charges and the dipoles of the D1 and D5 branes. Again this sector was
already studied in~\cite{Bena:2011dd} and it turns out that we can use
the same\footnote{We find it more convenient to work with gauge potentials rather than with field strengths, as instead was done in~\cite{Bena:2011dd}. Moreover the RR 3-form field strength used here should be identified with twice the 3-form $G$ of~\cite{Bena:2011dd}: this has the consequence that $2\,\Theta_I^{there}=\Theta_I^{here}$ and also that $2\, \hat{\psi}^{there}=\psi^{here}$. Moreover:
$2\,\gamma_2^{there} = d(\gamma_2^{here}+a_1\wedge \beta)$.}  set of equations also in our case. In order to put the D1 and
the D5 branes on the same footing, let us suppose that the gauge
potential $C^{(6)}$ takes a form which closely follows that of
$C^{(2)}$ in~\eqref{RRa}
\begin{equation}
  \label{eq:C6T4}
  C^{(6)} =  \left[-\frac{\alpha}{Z_2}\,(du+\omega)\wedge (dv+\beta) +
  a_2\wedge (dv+\beta) + \gamma_1\right] \wedge dV_{T^4} + \ldots
\,,
\end{equation}
where the dots stand for terms that do not have components along the
$T^4$. The equations we give below ensure that it is possible to
define a 1-form $a_2$ and a 2-form $\gamma_1$ so as to
satisfy~\eqref{eq:C6T4}. The 2-forms $\gamma_I$ satisfy
 \be\label{gammaeq}
D\gamma_2 = \star_4 (D Z_2 + \dot{\beta}\,Z_2)+ a_1\wedge D\beta \,,~~
D\gamma_1 = \star_4 (D Z_1 + \dot{\beta}\,Z_1)+ a_2\wedge D\beta \,. 
\ee
 Then it is convenient to combine $a_1$ and $a_2$ in two new 2-forms
 $\Theta_1$ and $\Theta_2$ 
\be
 \label{deftheta1}
 \Theta_1 = d a_1 + \partial_v (\gamma_2 - \beta\wedge a_1)~,~~~
 \Theta_2 = d a_2 + \partial_v (\gamma_1 - \beta\wedge a_2)\,,
\ee
which must satisfy the following duality conditions involving the
2-form $\psi$ defined in~\eqref{psidef}
 \be
 \label{selftheta1}
\star_4 (\Theta_1- Z_2 \,\psi) = \Theta_1 -Z_2\,\psi~,~~~
\star_4 (\Theta_2- Z_1 \,\psi) = \Theta_2 -Z_1\,\psi\,.
 \ee
The equations for $Z_1$ and $Z_2$ are a consequence of~\eqref{gammaeq}
and~\eqref{deftheta1}
\begin{equation}
  \label{eq:Zeq}
   D \star_4 (DZ_2 + \dot{\beta}\,Z_2) = - \Theta_1\wedge D\beta\,,\quad  
   D \star_4 (DZ_1 + \dot{\beta}\,Z_1) = - \Theta_2\wedge D\beta\,.
\end{equation}

Now we turn our attention to the equations that are sensitive to the
novelty of the ansatz considered in this paper. We will give only the
linearized version of these equations. We first have as set of
constraints which are the (linearized) analogue of~\eqref{gammaeq} and
\eqref{selftheta1}
\begin{equation}
  \label{eq:Z4eqlin}
  d \delta_2 = *_4 dZ_4~,~~~ 
  *_4 (da_4- \dot{\delta}_2) = da_4- \dot{\delta}_2~.
\end{equation}
There is also a constraint for the new component $x_3$ of the 4-form
gauge potential
\begin{equation}
  \label{eq:x3eqlin}
    dx_3 = *_4 \dot{Z}_4~. 
\end{equation}
Then we have the relation that constrains the angular momentum 1-form $\omega$, that can be derived, for example, by requiring the existence of the RR 6-form $C^{(6)}$. At the linearized level the new fields $Z_4, a_4, \delta_2, x_3$ do not enter this relation, and we can thus read it off from~\cite{Bena:2011dd}:
\begin{equation}
  \label{eq:omegaeqlin}
   d\omega+*_4 d\omega = (\Theta_1-\psi)+(\Theta_2-\psi) \,.
\end{equation}
This concludes the conditions following from supersymmetry. They also
imply all the second order equations of motion, except for the
$vv$-component of the Einstein equations. At the linearized order even this
extra constraint does not get modified by the new fields, and it reads
\begin{equation}
  \label{eq:Rvvlin}
  *_4 d *_4 \left( \dot{\omega} -\frac{1}{2}\,d\mathcal{F}\right) =\partial_v^2 (Z_1+Z_2)+
  \frac{1}{2}\,\partial_v^2 (h_{ii})\,.
\end{equation}

\section{Mixed disk amplitudes revisited} \label{sec:disc-rev}

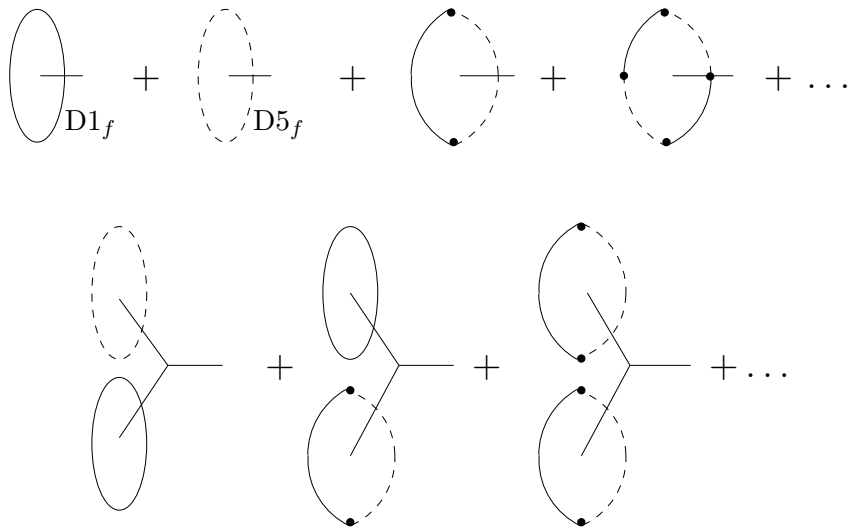
\begin{figure}[ht]
\begin{center}
\begin{picture}(0,0)%
\includegraphics{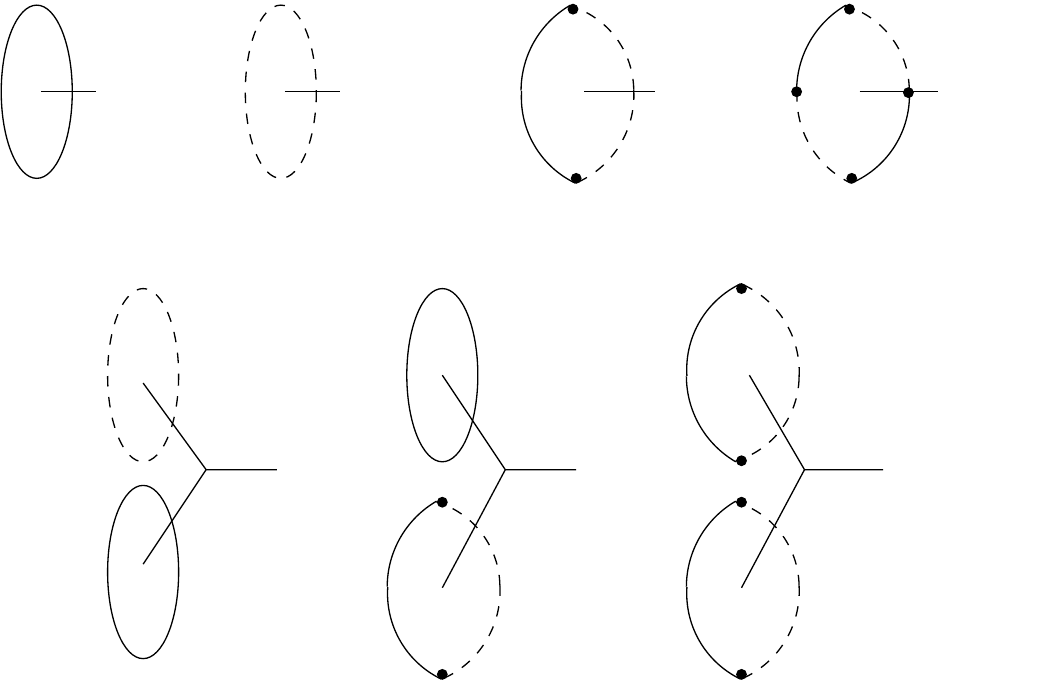}%
\end{picture}%
\setlength{\unitlength}{3315sp}%
\begingroup\makeatletter\ifx\SetFigFont\undefined%
\gdef\SetFigFont#1#2#3#4#5{%
  \reset@font\fontsize{#1}{#2pt}%
  \fontfamily{#3}\fontseries{#4}\fontshape{#5}%
  \selectfont}%
\fi\endgroup%
\begin{picture}(5919,3872)(262,-3606)
\put(2161,-2491){\makebox(0,0)[lb]{\smash{{\SetFigFont{14}{16.8}{\familydefault}{\mddefault}{\updefault}{\color[rgb]{0,0,0}+}%
}}}}
\put(3691,-2491){\makebox(0,0)[lb]{\smash{{\SetFigFont{14}{16.8}{\familydefault}{\mddefault}{\updefault}{\color[rgb]{0,0,0}+}%
}}}}
\put(5446,-2491){\makebox(0,0)[lb]{\smash{{\SetFigFont{14}{16.8}{\familydefault}{\mddefault}{\updefault}{\color[rgb]{0,0,0}+}%
}}}}
\put(5716,-2491){\makebox(0,0)[lb]{\smash{{\SetFigFont{14}{16.8}{\familydefault}{\mddefault}{\updefault}{\color[rgb]{0,0,0}\ldots}%
}}}}
\put(2071,-646){\makebox(0,0)[lb]{\smash{{\SetFigFont{11}{13.2}{\familydefault}{\mddefault}{\updefault}{\color[rgb]{0,0,0}D5$_f$}%
}}}}
\put(676,-646){\makebox(0,0)[lb]{\smash{{\SetFigFont{11}{13.2}{\familydefault}{\mddefault}{\updefault}{\color[rgb]{0,0,0}D1$_f$}%
}}}}
\put(1171,-331){\makebox(0,0)[lb]{\smash{{\SetFigFont{14}{16.8}{\familydefault}{\mddefault}{\updefault}{\color[rgb]{0,0,0}+}%
}}}}
\put(2701,-331){\makebox(0,0)[lb]{\smash{{\SetFigFont{14}{16.8}{\familydefault}{\mddefault}{\updefault}{\color[rgb]{0,0,0}+}%
}}}}
\put(4186,-331){\makebox(0,0)[lb]{\smash{{\SetFigFont{14}{16.8}{\familydefault}{\mddefault}{\updefault}{\color[rgb]{0,0,0}+}%
}}}}
\put(6166,-331){\makebox(0,0)[lb]{\smash{{\SetFigFont{14}{16.8}{\familydefault}{\mddefault}{\updefault}{\color[rgb]{0,0,0}\ldots}%
}}}}
\put(5851,-331){\makebox(0,0)[lb]{\smash{{\SetFigFont{14}{16.8}{\familydefault}{\mddefault}{\updefault}{\color[rgb]{0,0,0}+}%
}}}}
\end{picture}%
\end{center}
\caption{\label{Dpfig} \em A schematic picture of the diagrams
  relevant to the calculation of the gravitational backreaction of a
  microstate: the straight lines represent generic massless closed
  string states and the circles represent the boundary of the string
  world-sheet ending on the D-branes defining the microstate; the
  continuous (dashed) parts of the cricles mean that the boundary
  conditions appropriate for a vibrating D1-brane (D5-brane) are
  imposed on the string coordinates, while the black dots on the
  transition between these two types of boundary conditions represent
  the insertion of a twisted open string state (i.e. an open string
  stretched between a D1 and D5 brane). The first line contains all
  diagrams with a single border, i.e. the contributions linear in the
  source. The second line contains the diagrams that are needed to
  reconstruct the non-linear solution; the vertices in the bulk follow
  from the non-linear part of the standard IIB supergravity
  equations.}
\end{figure}

Let us start from an unbound point-like state whose basic
constituents are D1 branes wrapped on the $S^1$ and D5's wrapped on
the whole compact space. All D-branes vibrate in the non-compact space
according to the same profile $f_i(v)$. From the WCFT point of view
these D-branes can be described by using the boundary state formalism
(see~\cite{DiVecchia:1999rh,DiVecchia:1999fx} for a review), as
discussed
in~\cite{Hikida:2003bq,Blum:2003if,Bachas:2003sj}. In~\cite{Black:2010uq}
this approach was used to show that the boundary state $\ket{B}_{f^i}$
contains the information necessary to reconstruct the two-charge
solutions discussed in~\cite{Callan:1995hn,Dabholkar:1995nc} (once
they are rewritten in the appropriate duality frame). By following the
idea sketched in the Introduction, one can calculate the scalar
product of $\ket{B}_{f^i}$ with the various massless closed string
states. This gives the value of the one point correlators on a disk
where the boundary conditions are determined by $f^i(v)$. As shown
in~\cite{Black:2010uq}, these couplings can be combined with a free
propagator yielding the first two diagrams in Figure~\ref{Dpfig};
after a Fourier transformation from momentum to configuration space,
these diagrams reproduce the solution
in~\cite{Callan:1995hn,Dabholkar:1995nc} at the linear level in the D1
and D5 charges.

We are now interested in considering more in detail the zero-mode
structure of the boundary state, see Eqs.~(3.14) and (3.15)
of~\cite{Black:2010uq}. It follows that, even if both $t$ and $y$ are
directions with Neumann boundary conditions, D-branes with a
travelling wave can emit closed strings with a non-trivial momentum
$k$ along $v$ provided that
\begin{equation}
  \label{eq:pvpi}
  k_u = 0~,~~~~k_v + \dot{f}^i k_i =0\;,
\end{equation}
where, as before, the dot indicates the derivative with the respect to
$v$. Thus $k_v=0$ is possible for special values of the $k_i$'s. If we
limit ourselves to probes with zero momentum along $v$, then the
string correlators contain always an integral over $v$ and the smeared
solution discussed in~\cite{Black:2010uq} is recovered. However if we
test the D-brane configuration with a generic (localized) probe, then
we obtain a $v$-dependent result for the string correlator and the
original solution~\cite{Callan:1995hn,Dabholkar:1995nc}, without
integrals over $v$, is obtained. Notice that these $v$-dependent
2-charge solutions cannot be dualized to the D1-D5 frame as it was
done in~\cite{Lunin:2001fv}. The obstruction is clearly that for
$v$-dependent geometries the shifts along $y$ are not an isometry. This suggests
that the only 2-charge microstates in the D1-D5 frame are those
studied in~\cite{Lunin:2001fv} which always include a smearing over the $S^1$.

It is possible to reach the same conclusion by working directly in the
D1-D5 frame and following the microscopic approach used in this
paper. In this language the D1-D5 microstates are built by starting
from an unbound set of D-branes and switching on a vev for the open
strings stretched between the D1 and the D5 branes. If we do not want
to introduce any further charge or equivalently we wish to preserve
the same number of supercharges of the unbound system, then all open
string states introduced must have exactly zero momentum.  Now it is
clear also from this point of view why 2-charge configurations are
always smeared along the $S^1$: the boundary conditions appropriate
for the basic D-brane constituents require momentum conservation both
along $v$ and $u$, and no open strings carry any momentum; thus the
1-point correlators are non-trivial only if the closed string probe is
at zero momentum as well and then the results automatically include
integrals over both common Neumann directions.

We now wish to use localized probes to test the three charge systems
studied in~\cite{Giusto:2011fy}: this means that we start from the
unbound system mentioned at the beginning of this section, describing 
D1 and D5 branes oscillating with a common profile, and
introduce a vev for the open strings stretched between the D1 and the
D5 branes; then we probe the configuration with generic closed string
states which have also a non-zero momentum $k_v$. As
in~\cite{Giusto:2011fy}, in this section we limit ourselves to the
contributions of the first three diagrams in Figure~\ref{Dpfig} and
calculate explicitly the corresponding string amplitudes by using the
RNS formalism. In particular the simplest class of
microstates~\cite{Lunin:2001fv} corresponds to the configurations obtained by
introducing a vev for the mixed D1-D5 open strings in the Ramond
sector~\cite{Giusto:2009qq}. At the leading order, this condensate
involves only two open string states (the black dots in
Figure~\ref{Dpfig}) and so is described by a spinor bilinear which can
be decomponsed in a vector and a self-dual 3-form $v_{IJK}$ living in
the space orthogonal to the $T^4$. We will focus exclusively on the
contribution of the 3-form and, as done in~\cite{Giusto:2011fy}, we
also set to zero the components with just one (or all three) legs along the
$R^4$. So the non trivial part of $v_{IJK}$ can be decomposed in two
$SO(1,1) \times SO(4)$ rapresentations with opposite duality
properties
\begin{equation}
  \label{eq:vdualc}
  v_{uij} = -\frac{1}{2}\epsilon_{ijkl}\, v_{ukl}\,,\qquad 
  v_{vij} = +\frac{1}{2}\epsilon_{ijkl}\, v_{vkl}\; .
\end{equation}

The first guess is that the result will follow the same pattern
discussed above in the D1-P (or D5-P) case and that the $v$-dependent
backreaction would simply be the solution in Eq.~(5.16)--(5.32)
of~\cite{Giusto:2011fy}, where all integrals over $v$ (hidden in the
definition of ${\cal I}$) are dropped. However it is not difficult to
check that this guess cannot be correct, as the configuration just
mentioned is not a solution of the supergravity equations even if we
limit ourselves to the leading order in the charges and the
condensate~\eqref{eq:vdualc}. It turns out that in the 3-charge case
there are new contributions to the geometric backreaction that are
invisible to delocalized probes. The origin of this is as follows: the
string correlator is calculated in ten dimensions where the $R^4$ and
the light-cone $(u,v)$ directions share the feature of having the same
boundary conditions (either Neumann or Dirichlet) on both types of
branes. Then the correlators are more easily calculated in a $SO(1,5)$
invariant way which keeps all these directions on the same
footing. For instance this was done in Eq.~(4.14)
of~\cite{Giusto:2011fy} for a generic NS-NS probe. At this level the
generalization from a smeared to a localized probe involves just
dropping the integral over $v$. However, the ansatz of the previous
section is clearly not $SO(1,5)$ covariant; then in order to identify
the different supergravity fields we need to decompose the string
result in $SO(1,1)\times SO(4)$ representations. In doing this step
in~\cite{Giusto:2011fy} it was assumed that the momentum of the closed
string probe was entirely in the non-compact dimensions. So in order to
read the new $v$-dependent geometry we have to reconsider this
step. The starting point is Eq.~(4.14) of that paper which describes
the emission of NS-NS state from a disk with two twisted open string
state (i.e. the third diagram of the first line of Figure~\ref{Dpfig}
in the NS-NS sector). By dropping the $v$ integral we
have
\begin{equation}
  \label{eq:4.14}
  {\cal A}_{\rm NS}^{\rm D1D5} = - 2 \sqrt{2}\pi\, V_u\, e^{-ik_i f^i(v)}
  k^K {\cal G}^{IJ} ({}^{\rm t}R)_J^{~M} v_{IMK}~,
\end{equation}
where $V_u$ is the infinite volume of the $u$ direction, the uppercase
indices run over $v,u, x^i$ and $R$ is (the zero-mode part of) the
reflection matrix~\cite{Black:2010uq,Giusto:2011fy}\footnote{In~\cite{Black:2010uq,Giusto:2011fy}
the coordinates $u$ and $v$ where defined as in~\eqref{lc-coord} but
without the factor $1/\sqrt{2}$. In the following we adapt the results of those papers
to our conventions for the light cone coordinates: this is the reason why the form of the reflection matrices in~\eqref{eq:D1_bcs} and~\eqref{Rspinrep} differ from the corresponding expressions in~\cite{Giusto:2011fy}. In the expressions~\eqref{NSamplitude} and~\eqref{RRamplitude} for the NS-NS and RR amplitudes the metric component $\eta^{uv}$ equals $-1$ in the conventions of the present paper and $-2$ in the conventions of~\cite{Black:2010uq,Giusto:2011fy}.}
\begin{equation}
\label{eq:D1_bcs}
R^{\m}_{\phantom{\m \!}\n} ~=~ 
\left( \begin{array}{cccc} 
1 &   &  0  &  0 \\
 2 |\dot{f}({v})|^2 & 1 & -2 \dot{f}_i({v}) & 0 \\
2 \dot{f}_i({v}) &  0 & \!\! - \one  &  0  \\
  0  &  0  &  0 &  \!\! - \one
\end{array} \right) \;.
\end{equation}
The matrix indices are ordered by putting first the light-cone
coordinates $v,u$, then the $R^4$ indices $i$ and finally the $T^4$
ones $a$. By decomposing~\eqref{eq:4.14} we obtain
\bea
\label{NSamplitude}
\mathcal{A}_{NS}^{\rm D1D5} &=& 2\sqrt{2} \pi\, V_u \,e^{-ik_i f^i(v)}\,
k^l\,\Bigl[(\mathcal{G}^{uj}+ \mathcal{G}^{ju})\,v_{ujl}
+(\mathcal{G}^{vj}+ \mathcal{G}^{jv})\,v_{vjl}
\nonumber\\
&-& 2\,\eta^{uv}\,\mathcal{G}^{jv}\,|\dot{f}|^2\, v_{ujl} -2\,
\mathcal{G}^{uv}\,\dot{f}_j \,v_{ujl} -
2\,\mathcal{G}^{vv}\,\dot{f}_j\,v_{vjl}
+2\,\eta^{uv}\,\mathcal{G}^{ij}\,\dot{f}_j\,v_{uil} \Bigr]
\nonumber\\
&+& \eta^{uv}\,k_v \,\Bigl[-2\,\mathcal{G}^{jv}\,\dot{f}_i \,v_{uji} +
\mathcal{G}^{ij}\,v_{uij}\Bigr]\,, 
\label{genNSNS}
\eea
where ${\cal G}$ is the polarization of a generic NS-NS state which
needs to be further decomposed in the graviton, dilaton and
B-field. If we set $k_v$ to zero in~\eqref{genNSNS} of course we
recover Eq.~(4.15) of~\cite{Giusto:2011fy}. 

A similar step has to be performed also when we use a closed string
state in the RR sector as a probe. We start from Eq.~(4.24)
of~\cite{Giusto:2011fy}, again without the integral over $v$
\begin{equation}
  \label{eq:4.24}
  {\cal A}_R^{\rm D1D5} ~=~ \frac{i \pi}{2} V_u \,e^{- i k \cdot f
    (v) }\, v_{IJK} (C\Gamma^{IJK})^{BA}\, (C^{-1} {\mathbf F} {\cal R}
  C^{-1})_{AB}~, 
\end{equation}
where $A,B$ and $\Gamma$'s are spinor indices and the Gamma matrices
of $SO(1,5)$, $ {\mathbf F}$ is a bispinor encoding the RR field
strengths (see Eq.(4.23) of~\cite{Giusto:2011fy} for our conventions)
and ${\cal R}$ is the spinorial representation of the reflection
matrix, i.e.
\be \label{Rspinrep}
{\cal R} = \Gamma^{uv} - \dot{f}^i(v)\Gamma^{iv} \,.
\ee
Then by rewriting the field strengths in terms of the gauge potentials
$C$ we have
\bea
\label{RRamplitude}
\mathcal{A}_{RR}^{\rm D1D5} &=& 4\pi V_u
\,e^{- i k \cdot f(v)}\,\Bigl[\eta^{uv}\,k^l\,C^{(0)}\,\dot{f}_j\,v_{ulj}+\eta^{uv}\,
k^l\,C^{5678}\,\dot{f}_j\,v_{ulj}+ k^i\,C^{uvjk}\,\dot{f}_k\,v_{uij}
\nonumber\\
&+& \frac{1}{2}\,k^k\,C^{uvij}\,\dot{f}_k \,v_{uij}
+\frac{\eta^{uv}}{2}\,k_v\,C^{vijk}\,\dot{f}_k\,v_{uij}+\frac{\eta^{uv}}{2}\,k_v\,C^{ij}\,v_{uij}-k^i\,C^{uj}\,v_{uij}
\nonumber\\
&+& k^i\,C^{vj}\,v_{vij}
- \eta^{uv}\,k_v\,C^{vj}\,\dot{f}_l\,v_{ulj}
-k^i\,C^{uv}\,\dot{f}_l\,v_{uli}+\frac{\eta^{uv}}{2}\,k^l\,C^{ij}\,\dot{f}_l\,v_{uij}
\nonumber\\
&+&\eta^{uv}\,k^j\,C^{li}\,\dot{f}_l\,v_{uij}\Bigr]\,. 
\label{genRR}
\eea
Notice the appearance, in the second term of the second line, of a
contribution to the 4-form potential with three indices in the
$R^4$. Such a structure is absent in the smeared case where $k_v$ is
set to zero; this is the origin of the 3-form $x_3$ in the RR
ansatz~\eqref{RRa}. 

In our discussion so far we assumed that both the D1 and the D5 branes
are wrapped once on the $S^1$ of radius $R$; in this case the
functions $f_i(v)$ describing their (common) shape in the $R^4$ are
periodic under shifts $v \to v+2\pi R$. However the most interesting
configurations in the analysis of the black hole microstates involve
D-branes with wrapping number $w$ larger than one; then the profile
$f_i$ depends on the world-volume coordinate $\hat{v}$ and has
periodicity $2\pi w R$. Geometrically we can describe this situation
by splitting the closed profile in $w$ open segments $f^\a_i(v)$ with
$\a=1,\ldots,w$, and then imposing the gluing conditions $f^\a_i(2\pi R)
= f^{\a+1}_i(0)$ where $\a=w+1$ is identified with $\a=1$. We will not
write the complete expression describing the boundary state
corresponding to these multiply wrapped D-branes: since we consider
the emission only of closed string states with zero winding number, we
will treat each segment independently and sum over the individual
results to obtain the coupling of the wrapped D-brane to the closed
strings. This is sufficient for our purposes, since on the gravity
side we work at the linearized level in the sources, i.e. we are
interested in worldsheets with just one boundary. Another interesting
issue that we will not analyze further is related to the special
points in space-time where two of these segments intersect. The open
strings living at these intersections will feel locally two D-branes
with a relative rotation and boost; however the parameters of the
transformations are tuned so as to always preserve superysmmetry
(notice that the situation is different in the
space-like~\cite{Berkooz:1996km} or time-like~\cite{Bachas:1995kx}
cases). We will not make explicit use of these open string sectors. As
a consistency check of our approach, we will show that the
backreaction obtained by superimposing the contributions of each
segment preserves four supercharges.

It is now straightforward to follow the procedure discussed
in Section~4.3 of~\cite{Giusto:2011fy} and derive from \eqref{genNSNS}
and~\eqref{genRR} the configuration space
results~\eqref{stringfirst}--\eqref{basemetric}. As mentioned above,
the label $\a$ indicates the different segments of the multiply wrapped
profile (common to the D1 and D5 branes); also, in the expressions
below, a sum over $\a$, in each $\a$ dependent term, is present even
if it is not explicitly written
\begin{eqnarray}
\label{stringfirst}
Z_1 \!\!& =&\!\!  1+ Q_1 \,\mathcal{I}^{\a} + \mathbf{v}_{ulk}\, \partial_l \mathcal{I}^{\a}\dot{f}_k^\a\,,\\
Z_2 \!\!&=&\!\! 1+ Q_5 \,\mathcal{I}^{\a} + \mathbf{v}_{ulk}\, \partial_l \mathcal{I}^{\a}\dot{f}_k^\a\,,\\
\mathcal{F} \!\!&=&\!\! = -(Q_1+Q_5)\,\mathcal{I}^{\a}\,|\dot{f}^\a|^2-2\,\mathbf{v}_{vlk}\,\partial_l \mathcal{I}^{\a}\,\dot{f}_k^\a \,,\\
Z_4 \!\!&=\!\!& -\mathbf{v}_{ulk}\,\partial_l \mathcal{I}^{\a}\dot{f}_k^\a\,,\\
a_1 \!\!&=&\!\!  Q_5\,(\mathcal{I}^{\a} \dot{f}_i^\a + \dot{f}_k^\a
\widehat{\mathcal{I}}^{\a}_{ki})\,dx^i +\mathbf{v}_{uli}\,\partial_l
\mathcal{I}^{\a} |\dot{f}^\a|^2\,dx^i \,,\\
\label{betav}
\beta\!\!&=&\!\!  \mathbf{v}_{uli}\,\partial_l \mathcal{I}^{\a}\,dx^i \,,\\
a_4\!\!&=&\!\! \Bigl[\mathbf{v}_{uil}\,\partial_l \mathcal{I}^{\a}|\dot{f}^\a|^2 + \mathbf{v}_{uil} \,\partial_v ({\cal I}^{\a}\dot{f}_l)  \Bigr]\,dx^i\,,\\
\label{omegav}
\omega\!\!&=&\!\!  \Bigl[(Q_1 + Q_5)\,\mathcal{I}^{\a}\,\dot{f}_i^\a +\mathbf{v}_{vli} \,\partial_l \mathcal{I}^{\a} + \mathbf{v}_{uli}\,\partial_l \mathcal{I}^{\a}|\dot{f}^\a|^2 + {\mathbf v}_{uli}\, \partial_v ({\cal I}^{\a}\dot{f}_l^\a) \Bigr] dx^i ,\quad\phantom{}\\
\delta_2 \!\!&=&\!\!  \Bigl[\mathbf{v}_{uli}\,\partial_l \mathcal{I}^{\a}\dot{f}_j^\a-\frac{1}{2} \mathbf{v}_{uij} \,\partial_v{\cal I}^{\a} \Bigr]\,dx^i\wedge dx^j\,,\\
\gamma_2 \!\!&=\!\!&  \frac{1}{2}\, Q_5\,\widehat{\mathcal{I}}^{\a}_{ij}\,dx^i\wedge dx^j- \mathbf{v}_{uli}\,\partial_l \mathcal{I}^{\a}\dot{f}_{j}^\a\,dx^i\wedge dx^j \,,\\
x_3\!\!&=\!\!& \frac{1}{2}\,\mathbf{v}_{uij}\,\partial_v(\mathcal{I}^{\a}\dot{f}_k^\a)\,dx^i\wedge dx^j\wedge dx^k\,,\\
ds^2_4\!\!&=&\!\! [\delta_{ij} + \mathbf{v}_{uli}\,\partial_l \mathcal{I}^{\a}\,\dot{f}_j^\a +\mathbf{v}_{ulj}\,\partial_l \mathcal{I}^{\a}\,\dot{f}_i^\a - \delta_{ij} \,\mathbf{v}_{ulk}\,\partial_l \mathcal{I}^{\a}\,\dot{f}_k^\a]\,dx^i dx^j\,,
\label{basemetric}
\end{eqnarray}
where the $Q_I$'s indicate the D1 and D5 charges and $\mathbf{v}$ is the
open string condensate~\eqref{eq:vdualc} after a constant rescaling
\be
{\mathbf v}_{uij} = -\frac{2 \sqrt{2} \kappa}{\pi V_4}\,v_{uij}\,,\quad {\mathbf v}_{vij} = -\frac{2 \sqrt{2} \kappa}{\pi V_4}\,v_{vij}\,.
\ee
The function ${\cal I}^{\a}$ is
harmonic and defines implicitly also the 2-form $\widehat{\mathcal{I}}^{\a}$
\begin{equation}
  \label{eq:calI}
  {\cal I}^{\a} = \frac{1}{|x-f^{\a}(v)|^2}~,~~~
  d\widehat{\mathcal{I}}^{\a} = *_4 d{\cal I}^{\a}~,
\end{equation}
where the star is defined by using the flat $R^4$ metric and the
differential $d$ acts only on $x^i$ and not on $v$. In the next
section we check that the IIB background defined by the data above
solves the linearized constraints following from supersymmetry and the
equations of motion.

\section{The linearized geometry} \label{sec:lin-geo}

The linearized type IIB background obtained in the previous section
has different 2-charge limits. We can switch off the condensate of
D1-D5 strings $\mathbf{v}$ and obtain the geometry corresponding to an
unbound configuration of D1 and D5 branes or set to zero the
geometric profile $f^\a_i(v)$ and obtain the two charge D1-D5
microstates. As mentioned before, these D1-D5 geometries are dual to
the geometry of a vibrating string and their non-linear completion is
known~\cite{Lunin:2001fv,Lunin:2001jy,Kanitscheider:2007wq}. In particular, the dependence of the full solution on the open
string condensates can be expressed in terms of auxiliary periodic
functions $g_i(v')$, whose moments are the vev's used in the world-sheet
description of the previous section. For instance the condensate we
considered (i.e. $\mathbf{v}$) is written in terms of $g_i(v')$ as
follows~\cite{Giusto:2009qq}
\begin{equation}
  \label{eq:v2g}
  \mathbf{v}_{tij} \sim \frac{1}{L'} \int_0^{L'} \dot{g}_i(v')
  g_j(v')~, 
\end{equation}
where $L'$ is the periodicity of $g_i(v')$ and the dot on $g_i$ represents
the derivative with respect to the auxiliary variable $v'$. Thus we
can use the exact 2-charge D1-D5 solution to generalize the
result~\eqref{stringfirst}--\eqref{basemetric}. The idea is to promote
all $Q_I$ and $\mathbf{v}$-dependent terms to more general objects
depending on $g_i(v')$ which should encode the exact-dependence on the
open string condensates. This should account for the contribution of
the diagrams that have more than two insertions of twisted open
strings, see for instance the diagram at the end of the first line of
Figure~\ref{Dpfig}.

We can implement the idea above by setting to zero the profile
$f^\a_i(v)$. Then from the Lunin-Mathur~\cite{Lunin:2001fv,Lunin:2001jy} solution we can read
the general dependence of the various object on the functions $g_i(v')$
representing general twisted open string condensates. $Z_1$ and $Z_2$,
which for $f^\a_i(v)=0$ are just harmonic functions centered in zero,
become
\bea
Z_1^{\scriptscriptstyle D1D5} = 1+ \frac{Q_5}{L}\,\int_0^L \frac{dv'\,|\dot{g}(v')|^2}{|x - g(v')|^2}\,,\quad Z_2^{\scriptscriptstyle D1D5} = 1+ \frac{Q_5}{L}\,\int_0^L \frac{dv'}{|x - g(v')|^2}\,.
\eea
Similarly the general form of~\eqref{betav} and~\eqref{omegav} (always
at $f_i^\a(v)=0$), is given by
\be
\beta^{\scriptscriptstyle D1D5} = A^{\scriptscriptstyle
  D1D5}-B^{\scriptscriptstyle D1D5}\,,\quad 
\omega^{\scriptscriptstyle D1D5} = A^{\scriptscriptstyle D1D5} +
B^{\scriptscriptstyle D1D5}  \,, 
\ee
with
\be
A^{\scriptscriptstyle D1D5} = -\frac{Q_5}{L}\,\int_0^L\frac{dv'\,\dot{g}_i(v')}{|x - g(v')|^2}\,dx^i\,,\quad *_4 dB^{\scriptscriptstyle D1D5} = - dA^{\scriptscriptstyle D1D5}\,.
\ee
These quantities satisfy simple harmonic conditions
\begin{equation}
  \label{eq:Zihar}
  d*_4 d {Z_1}^{\scriptscriptstyle D1D5}=0~,~~~
  d*_4 d {Z_2}^{\scriptscriptstyle D1D5}=0~,
\end{equation}
and the self-duality and anti-self-duality properties
\begin{equation}
  \label{eq:D1D5ha1}
  *_4 d \beta^{\scriptscriptstyle D1D5} =  d \beta^{\scriptscriptstyle D1D5}~,~~
  *_4 d \omega^{\scriptscriptstyle D1D5} = -d\omega^{\scriptscriptstyle D1D5}~.
\end{equation}
Thanks to~\eqref{eq:Zihar} we can define a 2-form $\gamma_2^{D1D5}$
satisfying
\begin{equation}
  \label{eq:gammaD1D5}
  d\gamma_2^{D1D5} = *_4 d Z_2^{D1D5}.
\end{equation}
Moreover, by possibly adding exact terms, we can also impose the gauge
conditions
\begin{equation}
  \label{eq:D1D5gc1}
     d*_4 \beta^{\scriptscriptstyle D1D5} =  d*_4
     \omega^{\scriptscriptstyle D1D5} = 0~,
\end{equation}
which are satisfied by the perturbative expressions of the previous
section when ${f}_i^\a(v)=0$. In this gauge it is possible to
define a 2-form $\zeta^{\scriptscriptstyle D1D5}$ such that
\be\label{defzeta}
d \zeta^{\scriptscriptstyle D1D5} = *_4
\beta^{\scriptscriptstyle D1D5}\,, 
\ee
where $\zeta^{\scriptscriptstyle D1D5}$ itself is defined up to a
gauge, which we can fix by imposing
\be\label{gaugezeta}
\zeta^{\scriptscriptstyle D1D5} = - *_4 \zeta^{\scriptscriptstyle
  D1D5}\,. 
\ee
Now the strategy is to include the dependence on the geometric profile $f^\a_i(v)$
as done in the previous section. First let us introduce the barred
quantities which are related to the D1-D5 ones as follows
\begin{eqnarray}
  \label{eq:baqua}
  \bar Z_I^\a = Z_I^{\scriptscriptstyle D1D5}(x-f^\a(v))~,&~~
   \bar\gamma^\a_2 = \gamma_2^{\scriptscriptstyle D1D5}(x-f^\a(v))~,\\ \nonumber
 \bar\beta^\a = \beta^{\scriptscriptstyle D1D5}(x-f^\a(v))\;,&~~ \bar\omega^\a = \omega^{\scriptscriptstyle D1D5}(x-f^\a(v))~,~~
  \bar\zeta^\a = \zeta^{\scriptscriptstyle D1D5}(x-f^\a(v))\;.
\end{eqnarray}
These new expressions still solve, of course, the same harmonic equations and
duality conditions of the $f$-independent results written above
\bea
  \label{eq:D1D5ha}
  &&d*_4 d \bar{Z}_I^\a=0~,~~ d\bar\gamma^\a_2= *_4 d \bar Z^\a_2~,~~
  *_4 d \bar\beta^\a =  d \bar\beta^\a~,~~
  *_4 d \bar\omega^\a = -d\bar\omega^\a~,\nonumber\\
  &&  d*_4 \bar\beta^\a=  d*_4 \bar\omega^\a=0~,~~ d\bar\zeta^\a=*_4\bar\beta^\a~,~~\bar\zeta^\a=-*_4\bar\zeta^\a~.
\eea
The $v$-dependence of the barred quantities is entirely implicit in their dependence on $f^\a_i(v)$, so that, for example, 
\be\label{eq:betadot}
\partial_v \bar\beta^\a = - \dot{f}^\a_i \partial_i \bar\beta^\a\,.
\ee 
We will make repeatedly use of this identity in the following.
%

When only the first non-trivial term in the small $g_i(v')$ expansion of
these results is kept, $\bar\beta^\a$ reduces to~\eqref{betav} 
\begin{equation}
  \label{eq:betaD1D5}
    \bar\beta^\a = \mathbf{v}_{uli}\,\partial_l \mathcal{I}^{\a}\,dx^i\,,~~~
\end{equation}
and the explicit expression for $\bar\zeta^\a$ is
\begin{equation}
  \label{eq:zetaD1D5}
  \bar \zeta^\a = -\frac{1}{2}\,\mathbf{v}_{uij}\,{\cal I}^\a\,dx^i\wedge dx^j~,
\end{equation}
which satisfies~\eqref{eq:D1D5ha} thanks to~\eqref{eq:vdualc}.
Then we can generalize~\eqref{stringfirst}--\eqref{basemetric} by
redefining all the expressions appearing there as follows
\begin{eqnarray}\label{gZ1}
Z_1 \!\!& =&\!\!  {\bar Z}_1^\a + \bar \beta_k^\a\,\dot{f}_k^\a\,,\\
Z_2 \!\!&=&\!\! {\bar Z}_2^\a+ \bar \beta_k^\a\,\dot{f}_k^\a\,,\\
\mathcal{F} \!\!&=&\!\! =-({\bar Z}_1^\a+{\bar Z}_2^\a-2)\,|\dot{f}^\a|^2-2\,{\bar \omega^\a}_k\,\dot{f}_k^\a \,,\\
\label{gZ4}
Z_4 \!\!&=\!\!& -\bar \beta_k^\a\,\dot{f}_k^\a\,,\\
a_1 \!\!&=&\!\!  ({\bar Z}_2^\a-1)\,\dot{f}_i^\a\,dx^i - {\bar \gamma}_{2\,ik}^\a\,\dot{f}_k^\a\,dx^i + \bar \beta^\a\,|\dot{f}^\a|^2\,,\\
\label{gbeta}
\beta\!\!&=&\!\!  \bar\beta^\a\,,\\
a_4\!\!&=&\!\! -\bar\beta^\a\,|\dot{f}^\a|^2 + \partial_v (\bar \zeta_{ki}^\a\,\dot{f}_k^\a\,dx^i)\,,\\
\omega\!\!&=&\!\! ({\bar Z}_1^\a + {\bar Z}_2^\a-2)\,\dot{f}_i^\a\,dx^i  +{\bar \omega^\a}+ \bar \beta^\a\,|\dot{f}^\a|^2 - \partial_v (\bar \zeta_{ki}^\a\,\dot{f}_k^\a\,dx^i)\,,\\
\label{gdelta2}
\delta_2 \!\!&=&\!\!  \bar \beta^\a\wedge \dot{f}_i^\a\,dx^i +\partial_v \bar \zeta^\a\,,\\
\label{ggamma2}
\gamma_2 \!\!&=\!\!&  {\bar \gamma}_2^\a - \bar \beta^\a\wedge \dot{f}_{i}^\a\,dx^i\,,\\
x_3\!\!&=\!\!& -\partial_v (\bar \zeta^\a\wedge \dot{f}_i^\a\,dx^i)\,,\\
\label{gbasemetric}
ds^2_4\!\!&\equiv&\!\! (\delta_{ij} + h^{(1)}_{ij})\,dx^i dx^j =(\delta_{ij} +\bar\beta_i^\a\,\dot{f}_j^\a +\bar\beta_j^\a\,\dot{f}_i^\a - \delta_{ij} \,\bar\beta_k^\a\,\dot{f}_k^\a)\,dx^i dx^j\,.
\end{eqnarray}
As before we understood a sum over the label $\a$ indicating the
contribution of each segment of the multiply wrapped D1 and D5
branes. The two-step procedure used to
derive~\eqref{gZ1}--\eqref{gbasemetric} seems justified from the
world-sheet picture, where the data of the two profiles $g_i(v')$ and
$f^\a_i(v)$ are encoded by completely different open string states. We will
show that this more general configuration satisfies the supergravity
equations just as a consequence
of~\eqref{eq:D1D5ha}. Of course this implies that
also the configuration of the previous section, where we kept only the
first order terms in the small $g_i(v')$ expansion, is a solution of the
supergravity equations.

We leave most of the details of the check to the Appendix~\ref{Appa}
and collect in the main text only some results on the ``almost
hyperkahler" base metric. The first ingredient on which the whole
solution is built is the 1-form $\beta$. Since we are working at the
linearized order in $\beta$ itself, Eq.~\eqref{bstarb} simplifies: the
curved star $\star_4$ reduces to the flat one $*_4$ and the
$v$-dependent differential $D$ becomes the standard differential $d$ in
$R^4$. Then the linearized~\eqref{bstarb} is just $d\beta=*_4d\beta$
and it is a direct consequence of~\eqref{eq:D1D5ha}. The next step is
to define the set of complex structures $J^{(A)}$ compatible with the
4D metric~\eqref{gbasemetric}. In our case, these can be written in
terms of $\bar\beta^\a$ and the trivial complex structures $J_0^{(A)}$
appropriate for a flat $R^4$
\begin{align}
  \label{eq:Jflattris}
  J^{(1)}_0 &= dx^1 \wedge dx^2 - dx^3 \wedge dx^4 ~,
\\ \nonumber
  J^{(2)}_0 &= dx^1 \wedge dx^3 + dx^2 \wedge dx^4 ~,
\\\nonumber
  J^{(3)}_0 &= dx^1 \wedge dx^4 - dx^2 \wedge dx^3 ~.
\end{align}
Then at linear order in $\beta$ we have
\begin{eqnarray}
  \label{eq:Jbeta}
J^{(A)} \equiv  J_0^{(A)} +  J_1^{(A)} & = & J_0^{(A)} - \frac{1}{2}
\left[\dot{f}_i^\a \left(\bar\beta^\a \wedge J^{(A)}_0\right)_{ijk}
  dx^j \wedge dx^k\right] 
\\ \nonumber & = &
J_0^{(A)} - \bar\beta_i^\a\dot{f}_i^\a J^{(A)}_0 - \bar\beta^\a\wedge
J^{(A)}_{0\, ij}\dot{f}_j^\a dx^i\,.
\end{eqnarray}
At first order in $\beta$ the constraints~\eqref{eq:Jeq} reduce to
\begin{equation}
  \label{eq:Jeqlin}
  J^{(A)}_0\wedge J^{(B)}_1 + J^{(A)}_1\wedge J^{(B)}_0 = - h^{(1)}_{kk} \, \delta^{AB}\,
  dx^1 \wedge dx^2 \wedge dx^3 \wedge dx^4\;,~~
  d J^{A}_1 = \dot{\beta}\wedge J^{A}_0\,,
\end{equation}
where the trace of the first order part of the metric is
\begin{equation}
  \label{eq:trh}
  h^{(1)}_{kk} = - 2\,\bar\beta_k^\a\,\dot{f}_k^\a\,. 
\end{equation}
We leave to the Appendix~\ref{Appa} the proof that~\eqref{eq:Jeqlin}
follows from~\eqref{eq:Jbeta}, \eqref{eq:Jflattris} and $d\beta = *_4 d\beta$. Let us
conclude the discussion of the 4D base by noticing that the
(linearized) $\psi$ takes a very simple form
\begin{equation}
  \label{eq:psibeta}
  \psi = -\frac{1}{2}\,\sum_C \partial_v (\bar\beta_i^\a\,\dot{f}_j^\a)
  \,J^{(C)}_{0\,ij}\,J^{(C)}_0 = -\frac{1}{2}\,\partial_v
  [\bar\beta^\a \wedge \dot{f}_i^\a\,dx^i - *_4 (\bar\beta^\a\wedge
  \dot{f}_i^\a\,dx^i ) ] \,,
\end{equation}
where, in order to get the second identity, we used
\begin{equation}
  \label{eq:J0^2}
  \sum_C J_{0\,ij}^{(C)} J_{0\,kl}^{(C)} = \delta_{ik} \delta_{jl} -
  \delta_{il} \delta_{jk} - \epsilon_{ijkl} ~.
\end{equation}

\section{A (partially) non-linear generalization} \label{sec:BGSW}

The results of the previous section contain all the genuinely stringy
information on the superstrata that can be built by giving a (common)
non-trivial profile to the D1 and D5 branes present in the
Lunin-Mathur two-charge microstates. In the language of perturbative
string amplitudes this information captures the direct couplings
between the supergravity fields and the D-branes forming the bound
state, as depicted in the first line of Figure~\ref{Dpfig}. In other
words, the string calculation gives an explicit expression for the
stress-energy side of Einstein's equations and the
``source-depending'' side of all other supergravity equations. Clearly
a string-theory derivation of the diagrams depicted in the second line
of Figure~\ref{Dpfig} is very challenging, as it requires to deal with
multi-loop open string diagrams (i.e. world-sheets with many
boundaries). It is certainly easier to ignore all $\alpha'$
corrections and derive these non-linear terms by using supergravity:
technically we have just to use the stringy results for the disk
amplitudes as boundary conditions at large distances which fix the
solution of the source-free supergravity equation.

At present, the only fully non-linear $v$-dependent solution
representing a three charge microstate with a known CFT dual is the
one discussed in~\cite{Ford:2006yb}. This example can be interpreted
as a special configuration where the functions $g_i(v')$ introduced in
the previous section are directly determined by the geometric profile
$f_i(v)$. The solution of~\cite{Ford:2006yb} thus depends on only one
independent profile and it represents a set of measure zero in the
space of 3-charge microstates, that are expected to be parametrized by
generic functions of two variables~\cite{Bena:2011uw}.  A first level
of generalization consists in finding $v$-dependent solutions where
the profiles $g_i(v')$ and $f_i(v)$ are unrelated. In this section we
make a first step, by focusing an the class of superstrata considered
in~\cite{Bena:2011dd}. In that paper an exact solution carrying D1, D5
and P charges was constructed: that solution represents an unbound
state of D1 and D5 branes oscillating according to a profile which has
periodicity $2\pi R$ even if the D-branes are multiply wrapped. In the
notations introduced in Section~\ref{sec:disc-rev}, we then have
$f^\alpha_i(v)=f_i(v)$ for any $\alpha$. The solution discussed
in~\cite{Bena:2011dd} is exact in the D1 and D5 charges and in the
corresponding dipole charges, originating from the oscillation of the
global D1 and D5 charges, but has no KK-monopole
charge. 

The aim of this section is to generalize the solution
of~\cite{Bena:2011dd}: we keep the full dependence on the D1 and D5
(dipole) charges, but we also wish to include at first order the
effects of the KK-monopole dipole charge and the angular momentum;
this will also turn the configuration into a real bound state. In the
diagrammatic language of Figure~\ref{Dpfig}, this solution captures
also some of the non-linear terms depicted in the second line: in
particular we need to include all diagrams with at most one boundary
which can contribute to $\beta$ and $\omega$ at the linear level, but
with an arbitrary number of other type of boundaries. However, as
mentioned above, in order to capture these contributions we will not
follow the perturbative approach, but a trick closely related to the
approach of~\cite{Garfinkle:1990jq}, which was also used
by~\cite{Callan:1995hn,Dabholkar:1995nc} to generate the gravity
solutions corresponding to an oscillating string. The key point is the
following: when $f^\alpha_i(v)=f_i(v)$, the dependence on $f_i(v)$ of
the 4D base metric $ds^2_4$ in~\eqref{gbasemetric} can be completely
absorbed in the change of coordinates $x^i \to x^i + f^i(v)$. This
suggests that, when the solution~\eqref{gZ1}--\eqref{gbasemetric}
depends on a common profile, it takes a particularly simple form in
the new coordinate system. The result obtained after the shift can
again be expressed in terms of the ansatz discussed in
Section~\ref{sec:ansatz} and we obtain the following simple set of
geometric data
\begin{eqnarray}\label{flatZ1}
Z_1' \!\!& =&\!\!  Z^{\scriptscriptstyle D1D5}_1 \,,\\
Z_2' \!\!&=&\!\! Z^{\scriptscriptstyle D1D5}_2 \,,\\
\mathcal{F}' \!\!&=&\!\! =-|\dot{f}|^2 - 2\,\zeta^{\scriptscriptstyle D1D5}_{lk}\,\ddot{f}_l\,\dot{f}_k\,,\\
Z_4' \!\!&=\!\!& -\beta^{\scriptscriptstyle D1D5}_k\,\dot{f}_k\,,\\
a_1' \!\!&=&\!\! -\dot{f}_i\,dx^i \,,\\
\beta'\!\!&=&\!\! \beta^{\scriptscriptstyle D1D5}\,,\\
a_4'\!\!&=&\!\!\zeta^{\scriptscriptstyle D1D5}_{ki}\,\ddot{f}_k\,dx^i\,,\\
\omega'\!\!&=&\!\! -\dot{f}_i\,dx^i +\omega^{\scriptscriptstyle D1D5}+ \beta^{\scriptscriptstyle D1D5}\,|\dot{f}|^2 -\beta^{\scriptscriptstyle D1D5}_k\,\dot{f}_k\,\dot{f}_i\,dx^i \nonumber\\
&&+ \partial_l\zeta^{\scriptscriptstyle D1D5}_{ki}\,\dot{f}_l\,\dot{f}_k\,dx^i - \zeta^{\scriptscriptstyle D1D5}_{ki}\,\ddot{f}_k\,dx^i \,\\
\delta'_2 \!\!&=&\!\!  \beta^{\scriptscriptstyle D1D5}\wedge \dot{f}_i\,dx^i - \partial_k\zeta^{\scriptscriptstyle D1D5}\,\dot{f}_k\,,\\
\gamma'_2 \!\!&=\!\!&  \gamma^{\scriptscriptstyle D1D5}_2 - \beta^{\scriptscriptstyle D1D5}\wedge \dot{f}_{i}\,dx^i\,,\\
x'_3\!\!&=\!\!& \partial_k\zeta^{\scriptscriptstyle D1D5}\,\dot{f}_k \wedge \dot{f}_i\,dx^i - \zeta^{\scriptscriptstyle D1D5}\wedge \ddot{f}_i\,dx^i \,,\\
ds^{'2}_4\!\!&=&\!\! dx^i\,dx^i\,.
\label{flatbasemetric}
\end{eqnarray}
In Appendix~\ref{Appb} we give the explicit expression of the relation
between the new and the old geometric data induced by an
$f_i(v)$-dependent shift of the coordinates for the $R^4$. Clearly the
supergravity configuration obtained in this way is guaranteed, by
construction, to solve the equations of motion in the same
approximation used in the previous section, i.e. at first order in a
simultaneous expansion in both the D1 and D5 charges and the
KK-monopole dipole charge (to which $\beta^{\scriptscriptstyle D1D5}$,
and therefore $\zeta^{\scriptscriptstyle D1D5}$, and $\omega^{\scriptscriptstyle D1D5}$ are proportional).

The situation is actually a bit better: a slightly modified ansatz
actually solves the supergravity equations {\em exactly} in the D1 and
D5 (dipole) charges, and at first order in the KK-monopole dipole
charge. The basic reason for this drastic simplification is twofold:
first in these coordinates the $R^4$ metric is
flat~\eqref{flatbasemetric}, and second the combinations $\Theta'_1$
and $\Theta'_2$ vanish
\be
\Theta'_1=\Theta'_2=0\,,
\ee
as can be easily verified by using the
definitions~\eqref{deftheta1}. For instance
\be
\Theta'_1 = \partial_v (\gamma'_2 - \beta'\wedge a'_1) = \partial_v (\gamma^{\scriptscriptstyle D1D5}_2 - \beta^{\scriptscriptstyle D1D5} \wedge \dot{f}_i\,dx^i + \beta^{\scriptscriptstyle D1D5}\wedge \dot{f}_i\,dx^i)=0\,,
\ee
which follows from the fact that $a'_1$ is constant and that all
quantities with the {\em D1D5} superscript are independent of $f_i$
and thus of $v$. Of course the presence of terms that do not vanish at
large $|x^i|$ in $a'_1$, $\omega'$ and ${\cal F}'$ means that this
solution is not asymptotically Minkowski and thus this
coordinate frame is not the most suited to study the physical
properties of the microstate geometry. However the frame where the
metric $ds^{' 2}_4$ is flat is the perfect setup to study the
non-linear corrections induced by the supergravity equations. So we
will use this approach as a way of generating non-linear solutions and
then transform them back with the opposite shift $x^i\to x^i-f^i(v)$
to asymptotically flat geometries which are directly relevant to the
problem of studying the three charge microstates.

Let us now discuss how~\eqref{flatZ1}--\eqref{flatbasemetric} need to
be modified in order to provide a solution at all orders in $Q_1$ and
$Q_5$, but only at the linearised level in $\beta^{\scriptscriptstyle
  D1D5}$ and $\omega^{\scriptscriptstyle
  D1D5}$, which capture the presence of a KK-monopole dipole charge. Actually, the only
equation\footnote{In principle also the equation~\eqref{eq:Rvvlin} for $\mathcal{F}$ receives non-linear corrections in the D1 and D5 charges, as can be seen for example from Eq.~(4.12) of~\cite{Bena:2011dd}. However, when $\Theta_I=0$, these corrections involve the $v$-derivatives of $Z_I$, and thus vanish for the ansatz~\eqref{flatZ1}--\eqref{flatbasemetric}.} that receives corrections at our level of approximation is
the one for $x'_3$. It can be shown~\cite{GMPR} that the
equation~\eqref{eq:x3eqlin} should be generalized as follows
\be
dx'_3 = d a'_4\wedge \gamma_2 - a'_1 \wedge d\delta'_2 + Z'_2 *_4
\,\partial_v Z'_4\,,
\ee
where we now have to consider the factor of $Z_2$ in the last term as
it contains the dependence on $Q_5$ which we wish to keep exact; also
we need to include the first two terms because, after the shift, $a_1$
is constant and the term $\gamma_2^{\scriptscriptstyle D1D5}$ in
$\gamma_2$ is independent of the KK-monopole dipole charge. The solution
of this equation, at linear order in the KK-monopole charge, is
\be
x'_3= \partial_k\zeta^{\scriptscriptstyle D1D5}\,\dot{f}_k \wedge \dot{f}_i\,dx^i - Z_2^{\scriptscriptstyle D1D5}\,\zeta^{\scriptscriptstyle D1D5}\wedge \ddot{f}_i\,dx^i + \zeta^{\scriptscriptstyle D1D5}_{ki}\,\ddot{f}_k\,dx^i\wedge \gamma^{\scriptscriptstyle D1D5}_2\,.\\
\ee
Thus summarizing, the configuration specified by the
data~\eqref{flatZ1}--\eqref{flatbasemetric}, where $x_3$ is
substituted with the result above, solves the supergravity equations at
linear order in $\beta^{\scriptscriptstyle D1D5}$ and $\omega^{\scriptscriptstyle D1D5}$ .

We leave the explicit check of this statement to a forthcoming
publication. Here we can support it by showing how the new solution
looks in the original frame, where the 10D metric is asymptotically
flat. Thus we can use the formulae of Appendix~\ref{Appb} and perform
the coordinate shift $x^i\to x^i-f^i(v)$, so as to go back to the
frame where the solution is asymptotically flat. However this time we
keep terms of all orders in $Q_1$ and $Q_5$ and linearize the change of
variables only in the KK-monopole charge. We thus arrive at the
solution specified by the following data
 \begin{eqnarray}
 \label{D1D5exactfirst}
Z_1 \!\!& =&\!\!  {\bar Z}_1(1 + \bar \beta_k\,\dot{f}_k)\,,\\
Z_2 \!\!&=&\!\! {\bar Z}_2 (1+ \bar \beta_k\,\dot{f}_k)\,,\\
\label{bgcalf}
\mathcal{F} \!\!&=&\!\! =-({\bar Z}_1{\bar Z}_2-1)(1+\bar \beta_k\,\dot{f}_k)\,|\dot{f}|^2-2\,{\bar \omega}_k\,\dot{f}_k \,,\\
\label{bZ4}
Z_4 \!\!&=\!\!& -\bar \beta_k\,\dot{f}_k\,,\\
a_1 \!\!&=&\!\!  ({\bar Z}_2-1)\,\dot{f}_i\,dx^i - {\bar \gamma}_{2 ik}\,\dot{f}_k\,dx^i +\bar Z_2\, \bar \beta\,|\dot{f}|^2\,,\\
\label{bbeta}
\beta\!\!&=&\!\!  \bar\beta\,,\\
a_4\!\!&=&\!\! -\bar\beta\,|\dot{f}|^2 + \partial_v (\bar
\zeta_{ki}\,\dot{f}_k\,dx^i)\,,\\
\label{bgomega}
\omega\!\!&=&\!\! ({\bar Z}_1{\bar Z}_2-1)\,(1+\bar \beta_k\,\dot{f}_k)\,\dot{f}_i\,dx^i  +{\bar \omega}+\bar Z_1 \bar Z_2\, \bar \beta\,|\dot{f}|^2 - \partial_v (\bar \zeta_{ki}\,\dot{f}_k\,dx^i)\,,\\
\delta_2 \!\!&=&\!\!  \bar \beta\wedge \dot{f}_i\,dx^i +\partial_v \bar \zeta\,,\\
\gamma_2 \!\!&=\!\!&  {\bar \gamma}_2 - \bar \beta\wedge (\dot{f}_{i}\,dx^i -\bar \gamma_{2\,ij}\,\dot{f}_i \,dx^j)\,,\\
x_3\!\!&=\!\!& -\partial_v \bar \zeta\wedge \dot{f}_i\,dx^i -\bar Z_2\,\bar \zeta \,\ddot{f}_i\, dx^i +(\bar \zeta_{ki}\,\ddot{f}_k\,dx^i + \bar \beta _k \,\dot{f}_k\,\dot{f}_i\,dx^i)\wedge \bar \gamma_2 \,,\\
\label{D1D5exactbase}
ds^2_4\!\!&=&\!\!  (\delta_{ij} +\bar\beta_i\,\dot{f}_j +\bar\beta_j\,\dot{f}_i - \delta_{ij} \,\bar\beta_k\,\dot{f}_k)\,dx^i dx^j\,,
\end{eqnarray}
where we used the same conventions of the previous section, but we
dropped all superscript $\alpha$, since we are now working under the
assumption $f^\alpha_i(v)=f_i(v)$. It is interesting to compare this
result with the solution of~\cite{Bena:2011dd}. 
Even if our solution falls into an enlarged ansatz, where all fields of type IIB
supergravity are non-trivial, the extra fields, which are encoded in $Z_4$, $a_4$, $\delta_2$ and $x_3$, arise from the combined effect of having both a KK-monopole charge and an oscillating
profile. Hence, when $\beta^{\scriptscriptstyle D1D5}$ and $\omega^{\scriptscriptstyle
D1D5}$ are set to zero our solution should reduce to the result of~\cite{Bena:2011dd}. This is indeed the case, as it can be
checked by comparing~\eqref{bgcalf} and~\eqref{bgomega} with
Eqs.~(4.11) and~(4.13) of~\cite{Bena:2011dd}, when the arbitrary
parameters ($c_1$, $c_2$ and the harmonic function $H_3$) in those equations are chosen appropriately. The geometric
data given above provide a generalization of the
result of~\cite{Bena:2011dd} which includes the first corrections in
$\beta^{\scriptscriptstyle D1D5}$ and $\omega^{\scriptscriptstyle
 D1D5}$.

\section{Conclusions} \label{sec:conclusions} In this paper we take
the first steps towards the construction of supergravity solutions
describing the class of bound states carrying D1, D5 and P charges
introduced in~\cite{Bena:2011uw} with the name of superstrata. The
example of superstrata we construct carry four dipole charges corresponding to
D1 and D5 branes, to an F1-string, and to a KK-monopole. We have
obtained the geometries via successive levels of approximation. First
we considered the solution as an expansion around flat space and for
the most part we discarded terms of order higher than the first in
this expansion. This corresponds to the
solution~\eqref{stringfirst}--\eqref{basemetric}, that results from
summing the first three types of string diagrams in
Figure~\ref{Dpfig}: it is a linearized solution in which, moreover,
the linear terms receive contributions only from a finite number (zero
or two) of insertions of the string condensate associated with the
open strings stretching between D1 and D5 branes.

Exploiting the fact that the D1-D5 solution, that resums arbitrary numbers of D1-D5 condensate insertions, is  known~\cite{Lunin:2001fv}, and that the dependence on the oscillation profile $f_i(v)$ can be exactly computed in the WCFT, we infer the geometry~\eqref{gZ1}--\eqref{gbasemetric}, that gives the complete linearized solution for a superstratum: this solution should contain the information of all the string disk diagrams and, together with the non-linear information encoded in the supergravity equations, should allow to reconstruct the full exact geometry. 

We make a first step towards the non-linear completion of the solution in the particular case in which all the strands of the multiply wound D1-D5 string are described by the same oscillation profile, i.e. when $f^\a_i(v)=f_i(v)\,,\,\forall \a$. In this case one can apply a trick analogous to the one used in~\cite{Garfinkle:1990jq,Callan:1995hn,Dabholkar:1995nc} and move to a coordinate frame where the equations simplify, though the solution ceases to be explicitly asymptotically Minkowski. Transforming back to an asymptotically flat frame, we arrive at the solution~\eqref{D1D5exactfirst}--\eqref{D1D5exactbase}, that solves the equations at all orders in the D1 and D5 charges, but only at first order in the KK-monopole dipole charge; this solution represents the first order deformation of the solution of~\cite{Bena:2011dd} upon the addition of the fourth dipole charge.

Our work opens the way to several future developments. The extension of our result to an exact
solution of supergravity will not only represent a technical improvement but it will provide important physical insights on the nature of black hole microstates: it will allow us to probe a larger class of three charge microstate geometries at scales where a classical horizon is expected to form, and to verify their smoothness or their eventual singularity structure. 

The solutions we find are $v$-dependent geometries that contain more fields (making up one more 6D tensor multiplet) than the ones present in the  
ansatz of~\cite{Bena:2011dd}. The conditions for supersymmetry in this enlarged $v$-dependent setting are not known: to aim at a non-linear extension of our results a first necessary step is thus the derivation of the appropriate set of supergravity equations. Work in this directions is in progress~\cite{GMPR}. 

With the supergravity equations at hand, and exploiting the trick introduced in Section~\ref{sec:BGSW}, we think that a fully non-linear completion of the solution~\eqref{D1D5exactfirst}--\eqref{D1D5exactbase}, describing a superstratum where the various strands oscillate with the same profile, should be within reach. 

For a generic superstratum, described by strands oscillating with independent profiles, the problem seems much more intricate, and potentially interesting\footnote{Already for solutions with no KK-monopole dipole charge it was noted in~\cite{Niehoff:2012wu} that new shape-shape interaction terms arise in generic superstrata with unequal strands.}. In particular there does not seem to exist a coordinate frame that trivializes the 4D base metric given in~\eqref{gbasemetric}. One is thus faced with the highly non-linear problem of finding an ``almost hyperkahler'' metric and a 1-form $\beta$ that solve the constraints~\eqref{bstarb}--\eqref{eq:Jeq} and reduce to~\eqref{gbasemetric} and~\eqref{gbeta} at the linear level. It was however shown in~\cite{Bena:2011dd} that this is the only intrinsically non-linear part of the problem: the remaining equations, if solved in the right order, reduce to a sequence of linear equations.

Finally we note that we landed onto a supergravity ansatz that
generalizes the one of~\cite{Bena:2011dd} by starting from the
simplest worldsheet string configuration describing a bound state of
D1-D5-P charges. In particular we decided to switch on only the 
components of the D1-D5 string condensate associated with the
2-charge microstates of~\cite{Lunin:2001fv}, but general condensates are possible, 
corresponding to the microstates of~\cite{Kanitscheider:2007wq}.
Moreover, we took the condensate to
be $v$-independent, so that momentum is entirely carried by the
oscillation profile $f_i(v)$. It is conceivable (and some preliminary
computations support this possibility) that by starting from a more
general worldsheet setup and by fine-tuning the various ingredients at
our disposal, one could engineer a microscopic worksheet configuration
that only sources the fields present in the restricted ansatz
of~\cite{Bena:2011dd}. Most likely, having a simpler supergravity
ansatz should contribute to make the task of constructing a fully
non-linear solution more tractable. The price to pay for this
simplification is that the microscopic D-brane configuration will be
more complicated and thus the derivation of the linearized solution
from string amplitudes will require more effort.

\noindent {\large \textbf{Acknowledgements} }

\vspace{2mm} We thank I.~Bena, G.~Dall'Agata, L.~Martucci, S.~Mathur, J.F.~Morales, M.~Petrini, M.~Shigemori, A.~Tomasiello, D.~Turton and N.~Warner 
for several enlightening discussions. SG has been partially supported by MIUR-PRIN contract 2009-KHZKRX, by the Padova University Project CPDA119349 and by INFN.
RR has been partially supported by STFC Standard Grant
ST/J000469/1 ``String Theory, Gauge Theory and Duality".

\appendix
\section{Checking the linearized equations of motion}
\label{Appa}

We will explicitly verify that the geometry given in~\eqref{gZ1}--\eqref{gbasemetric} solves the
linearized equations of motion as a consequence of~\eqref{eq:D1D5ha}.

We already noted in the text that $\beta$ trivially solves its equation \eqref{bstarb} at linear order. Let us now look at the equations for the 4D base $ds^2_4$: the linearized equations are given in~\eqref{eq:Jeqlin}. The first is an algebraic constraint that can be verified starting from the explicit form of $J^{(A)}_1$ given in~\eqref{eq:Jbeta}:
\bea
\frac{1}{2}(J^{(A)}_1\wedge J^{(B)}_0+J^{(B)}_1\wedge J^{(A)}_0) &\equiv & J^{((A)}_1\wedge J^{(B))}_0\nonumber\\
&=&  -\bar\beta^\a_i\,\dot{f}^\a_i \,J^{(A)}_0 \wedge J^{(B)}_0 - \bar\beta^\a\wedge J^{((A)}_{0\, ij}\,\dot{f}_j\,dx^i\wedge J^{(B))}_0\nonumber\\
&=& 2\,\bar\beta^\a_i\,\dot{f}^\a_i \,\delta^{AB}\,d^4x -\frac{1}{2}\,\epsilon_{ijkl}\,\bar\beta^\a_i\,J^{((A)}_{0\,jm}\,J^{(B))}_{0\,kl}\,\dot{f}^\a_l\,d^4x\nonumber\\
&=& 2\,\bar\beta^\a_i\,\dot{f}^\a_i \,\delta^{AB}\,d^4x + \bar\beta^\a_i \,J^{((A)}_{0\,jm}\,J^{(B))}_{0\,ij}\,\dot{f}^a_m\,d^4x\nonumber\\
&=&2\,\bar\beta^\a_i\,\dot{f}^\a_i \,\delta^{AB}\,d^4x - \bar\beta^\a_i \,\dot{f}^\a_i\,\delta^{AB}\,d^4x\nonumber\\
&=& \bar\beta^\a_i\,\dot{f}^\a_i \,\delta^{AB}\,d^4x = -\frac{1}{2}\,h_{kk}^{(1)}\,\delta^{AB}\,d^4x  \,.
\eea
Here we have introduced the short-hand notation
\be
d^4x\equiv dx^1\wedge dx^2\wedge dx^3 \wedge dx^4\,,
\ee
and we have used the anti-self-duality of $J^{(A)}_0$ and the property
\be\label{J0property}
J^{(A)}_{0\,ik}\,J^{(B)}_{0\,kj} = \epsilon^{ABC}\,J^{(C)}_{0\, ij}-\delta^{AB}\delta_{ij}\,,
\ee
which is the zeroth-order version of~\eqref{eq:acs}. The last step follows from~\eqref{eq:trh}.

The differential constraint in~\eqref{eq:Jeqlin} can be proved as follows:
\bea
dJ^{(A)}_1 - \dot{\beta}\wedge J^{(A)}_0 &=& -d (\bar\beta^\a_i\,\dot{f}^\a_i)\,J^{(A)}_0-d\bar\beta^\a\wedge J^{(A)}_{0\,ij}\,\dot{f}^\a_j\,dx^i + \dot{f}^\a_i\,\partial_i \bar\beta^\a \wedge J^{(A)}_0\nonumber\\
&=& dx^i\,\dot{f}^\a_j\wedge [(\partial_j\bar\beta^\a_i - \partial_i\bar\beta^\a_j)\,J^{(A)}_0-d\bar\beta^\a\,J^{(A)}_{0\,ij}]\nonumber\\
&=& -dx^i\,\dot{f}^\a_j\wedge [(d\bar\beta^\a)_{ij}\,J^{(A)}_0 + d\bar\beta^\a\,J^{(A)}_{0\,ij} ]\,,
\eea
where we used~\eqref{eq:betadot}.
To see that this is zero, let us take its Hodge dual
\be
*_4(dJ^{(A)}_1 - \dot{\beta}\wedge J^{(A)}_0) = -\frac{1}{2}\,dx^i\epsilon_{ijkl}\,[(d\bar\beta^\a)_{jm}\,J^{(A)}_{0\,kl}+ (d\bar\beta^\a)_{jk}\,J^{(A)}_{0\,lm}]\,\dot{f}^\a_m\,.
\ee
If we use the anti-self-duality of $J^{(A)}_0$ in the first term of the r.h.s. and the self-duality of $d\bar\beta^\a$ in the second term, we find
\be\label{eq:stardJ1}
*_4(dJ^{(A)}_1 - \dot{\beta}\wedge J^{(A)}_0) = dx^i\,[(d\bar\beta^\a)_{jm}\,J^{(A)}_{0\,ij}- (d\bar\beta^\a)_{il}\,J^{(A)}_{0\,lm}]\,\dot{f}^\a_m\,.
\ee
If, vice versa, we use the self-duality of $d\bar\beta^\a$ in the first term of the r.h.s. and the anti-self-duality of $J^{(A)}_0$ in the second term, we find
\bea\label{eq:stardJ2}
&&*_4(dJ^{(A)}_1 - \dot{\beta}\wedge J^{(A)}_0) = -\frac{1}{4}\,dx^i\epsilon_{ijkl}\,[\epsilon_{jmab}\,(d\bar\beta^\a)_{ab} J^{(A)}_{0\,kl} - (d\bar\beta^\a)_{jk}\,\epsilon_{lmab}\,J^{(A)}_{0\,ab}]\,\dot{f}^\a_m\nonumber\\
&&= \frac{1}{2}\,dx^i \,(d\bar\beta^\a)_{kl}\,J^{(A)}_{0\,kl}\,\dot{f}^\a_i +dx^i \,(d\bar\beta^\a)_{li}\,J^{(A)}_{0\,kl}\,\dot{f}^\a_k\nonumber\\
&&\qquad -\frac{1}{2}\,dx^i \,(d\bar\beta^\a)_{jk}\,J^{(A)}_{0\,jk}\,\dot{f}^\a_i-dx^i \,(d\bar\beta^\a)_{jk}\,J^{(A)}_{0\,ki}\,\dot{f}^\a_j\nonumber\\
&&=dx^i\,[(d\bar\beta^\a)_{il}\,J^{(A)}_{0\,lm} -(d\bar\beta^\a)_{jm}\,J^{(A)}_{0\,ij} ]\,\dot{f}^\a_m\,.
\eea
If we compare (\ref{eq:stardJ1}) and (\ref{eq:stardJ2}) we see that the expressions on the r.h.s. are equal and opposite and hence vanish: the second equation in~\eqref{eq:Jeqlin} is thus satisfied.

Let us now pass to the equation for $a_1$: at linear order the definition of $\Theta_1$ in~\eqref{deftheta1} becomes
\bea
\Theta_1 &=& da_1 + \dot{\gamma_2}\nonumber\\
&=&d({\bar Z}^\a_2 \,\dot{f}^\a_i\,dx^i - {\bar \gamma}^\a_{2\,ik}\,\dot{f}^\a_k\,dx^i) +\partial_v{\bar \gamma}^\a_2+ d\bar\beta^\a |\dot{f}^\a|^2 - \partial_v (\bar\beta^\a\wedge \dot{f}^\a_i\,dx^i)\,,
\eea
where the second expression follows from the ansatz~\eqref{gZ1}--\eqref{gbasemetric}. Using (the dual of) the identity 
\bea
*_4 d(- {\bar \gamma}^\a_{2\,ik}\,\dot{f}^\a_k\,dx^i) &=& -\frac{1}{2}\,dx^i\wedge dx^j\,\epsilon_{ijkl} \,\partial_k{\bar\gamma}^\a_{2\,lm}\,\dot{f}^\a_m \nonumber\\
&=&  -\frac{1}{4}\,dx^i\wedge dx^j\,\epsilon_{ijkl} \,\epsilon_{klmp}\,\partial_p {\bar Z}^\a_2\,\dot{f}^\a_m+\frac{1}{4}\,dx^i\wedge dx^j\,\epsilon_{ijkl}\, \partial_m{\bar\gamma}^\a_{2\,kl}\,\dot{f}^\a_m \nonumber\\
&=&d({\bar Z}^\a_2 \,\dot{f}^\a_i\,dx^i) -\partial_v (*_4 {\bar \gamma}^\a_2)\,,
\eea
that descends from the second relation in~\eqref{eq:D1D5ha}, we can rewrite
\be
\Theta_1=d({\bar Z}^\a_2 \,\dot{f}^\a_i\,dx^i ) + *_4 d({\bar Z}^\a_2 \,\dot{f}^\a_i\,dx^i) +d\bar\beta^\a |\dot{f}^\a|^2 -\partial_v (\bar\beta^\a\wedge \dot{f}^\a_i\,dx^i) \,.
\ee
The linearized version of the $a_1$ equation~\eqref{selftheta1} is
\be\label{eq:theta1lin}
*_4(\Theta_1-\psi) = \Theta_1-\psi\,.
\ee
Using the expression for $\psi$ given in~\eqref{eq:psibeta} and the one for $\Theta_1$ derived above one finds
\be\label{eq:theta1minuspsi}
\Theta_1-\psi = d({\bar Z}^\a_2 \,\dot{f}^\a_i\,dx^i ) + *_4 d({\bar Z}^\a_2 \,\dot{f}^\a_i\,dx^i) +d\bar\beta^\a |\dot{f}^\a|^2 -\frac{1}{2}\partial_v [\bar\beta^\a\wedge \dot{f}^\a_i\,dx^i + *_4 (\bar\beta^\a\wedge \dot{f}^\a_i\,dx^i)] \,,
\ee
which shows explicitly that $\Theta_1-\psi$ is self-dual, as required by~\eqref{eq:theta1lin}.

From the first of~\eqref{gammaeq}, the linearized version of the $Z_2$ equation is
\be\label{eq:gamma2lin}
d\gamma_2 = *_4 (dZ_2 + \dot{\beta})\,.
\ee
From the ansatz~\eqref{gZ1}--\eqref{gbasemetric}, and the identity~\eqref{eq:betadot}, one finds
\be
dZ_2 + \dot{\beta} = d{\bar Z}^\a_2 - dx^i\,(\partial_i\bar\beta^\a_k-\partial_k\bar\beta^\a_i)\,\dot{f}^\a_k\,;
\ee
hence, making use of the second and third relation in~\eqref{eq:D1D5ha}, one has
\bea
*_4(dZ_2 + \dot{\beta})&=& *_4 d{\bar Z}^\a_2 -\frac{1}{3!}\,dx^i\wedge dx^j\wedge dx^k\,\epsilon_{ijkl}\,(d\bar\beta^\a)_{lm}\,\dot{f}^\a_m\nonumber\\
&=&-d\bar\gamma^\a_2 - \frac{1}{2\,3!}\,dx^i\wedge dx^j\wedge dx^k\,\epsilon_{ijkl}\,\epsilon_{lmpq}\,(d\bar\beta^\a)_{pq}\,\dot{f}^\a_m\nonumber\\
&=& -d\bar\gamma^\a_2 +d\bar\beta^\a\wedge \dot{f}^\a_i\,dx^i = -d\gamma_2\,,
\eea
where in the last step we have used~\eqref{ggamma2}. We have thus obtained the Hodge dual of~\eqref{eq:gamma2lin}. As we have already explained, \eqref{eq:gamma2lin} implies the linearized version of the $Z_2$ equation in~\eqref{eq:Zeq}. 

Analogously, to prove the $Z_1$ equation in~\eqref{eq:Zeq} it is easier to show that there exists a $\gamma_1$ that solves the second equation in~\eqref{gammaeq}, which to linear order is
\be
d\gamma_1 = *_4 (dZ_1 + \dot{\beta})\,.
\ee
If one defines
\be\label{def:gamma1}
\gamma_1 = \bar\gamma^\a_1 -\bar\beta^\a\wedge \dot{f}^\a_i\,dx^i\,,
\ee
where
\be
d\bar\gamma^\a_1 = *_4 d\bar Z^\a_1\,,
\ee
(such a $\bar\gamma^\a_1$ exists thanks to the firts of~\eqref{eq:D1D5ha}) the proof proceeds as for $Z_2$.

Let us now come to the equations for the new multiplet, $Z_4, a_4, \delta_2, x_3$. To verify the first equation in~\eqref{eq:Z4eqlin}, let us start from~\eqref{gZ4} and compute
\bea
*_4 d Z_4 &=& -\frac{1}{3!}\,dx^i\wedge dx^j\wedge dx^k\,\epsilon_{ijkl}\,\partial_l\bar\beta^\a_m\,\dot{f}^\a_m\nonumber\\
&=& -\frac{1}{3!}\,dx^i\wedge dx^j\wedge dx^k\,\epsilon_{ijkl}\,(d\bar\beta^\a)_{lm}\,\dot{f}^\a_m-\frac{1}{3!}\,dx^i\wedge dx^j\wedge dx^k\,\epsilon_{ijkl}\,\partial_m\bar\beta^\a_{l}\,\dot{f}^\a_m\nonumber\\
&=& d\bar\beta^\a\wedge \dot{f}^\a_i\,dx^i+\partial_v(*_4\bar\beta^\a)= d(\bar\beta^\a\wedge \dot{f}^\a_i\,dx^i+ \partial_v\bar\zeta^\a)=d\delta_2\,,
\eea
where in the intermediate steps we have used again the self-duality of $d\bar\beta$ and the definition of $\bar\zeta$ in~\eqref{eq:D1D5ha} and the relation~\eqref{eq:betadot}, and in the last step we have compared with the form of $\delta_2$ given in~\eqref{gdelta2}.

To prove the $a_4$ equation (the second equation in~\eqref{eq:Z4eqlin}), we need the identity
\be\label{identityzeta}
*_4 d(\bar \zeta^\a_{ki}\,\dot{f}^\a_k\,dx^i) = \bar\beta^\a\wedge \dot{f}^\a_i\,dx^i + \partial_v\bar \zeta^\a\,,
\ee
which can be shown as follows
\bea
*_4 d(\bar \zeta^\a_{ki}\,\dot{f}^\a_k\,dx^i) &=& \frac{1}{2}dx^i\wedge dx^j\,\epsilon_{ijkl}\, \partial_k \bar \zeta^\a_{ml}\,\dot{f}^\a_m \nonumber\\
&=&\frac{1}{4}dx^i\wedge dx^j\,\epsilon_{ijkl} \,\epsilon_{kmlp}\,\bar\beta^\a_p\,\dot{f}^\a_m+\frac{1}{4}dx^i\wedge dx^j\,\epsilon_{ijkl} \,\partial_m\bar\zeta^\a_{kl}\,\dot{f}^\a_m\nonumber\\
&=&\bar\beta^\a\wedge \dot{f}^\a_i\,dx^i + \partial_v\bar\zeta^\a\,,
\eea
where we have used the defining properties of $\bar\zeta^\a$ (the last two identities in~\eqref{eq:D1D5ha}) and the analogue of~\eqref{eq:betadot} for $\bar\zeta^\a$. The ansatz~\eqref{gZ1}--\eqref{gbasemetric} gives
\be
 da_4 + \dot{\delta}_2 =  -d\bar\beta^\a\,|\dot{f}^\a|^2 + \partial_v [d(\bar \zeta^\a_{ki}\,\dot{f}^\a_k\,dx^i)+\bar\beta^\a\wedge \dot{f}^\a_i\,dx^i + \partial_v\bar \zeta^\a]\,;
 \ee
 the identity~\eqref{identityzeta} shows that the quantity in square brackets on the r.h.s. of the above expression is self-dual, while the self-duality of $d\bar\beta^\a$ guarantees that the first term on the r.h.s. is self-dual. Hence the second relation in~~\eqref{eq:Z4eqlin} is satisfied.
 
 To verify the linearized $x_3$ equation~\eqref{eq:x3eqlin} let us compute
 \be
 dx_3 =  -  \partial_v \,d(\bar\zeta^\a\wedge \dot{f}^\a_i\,dx^i) = - \frac{1}{2}\,\epsilon_{ijkl}\,\partial_v (\partial_i \bar\zeta^\a_{jk}\,\dot{f}^\a_l) \,d^4x = -\partial_v (\bar\beta^\a_i\,\dot{f}^\a_i)\,d^4x\,,
 \ee
where in the last step we have used that $d\bar\zeta^\a = *_4 \bar\beta^\a$; this is indeed equal to 
 \be
 *_4 \dot{Z}_4 = -*_4 \partial_v (\bar\beta^\a_i\,\dot{f}^\a_i)\,.
 \ee

Verifying the $\omega$ equation~\eqref{eq:omegaeqlin} amounts to show that the $\Theta_2$ derived from ~\eqref{eq:omegaeqlin} can be written as in second of~\eqref{deftheta1}, for some 1-form $a_2$. At the linear level one should have that
\be
\Theta_2 = d a_2 + \dot{\gamma_1}\,,
\ee
where $\gamma_1$ is given in~\eqref{def:gamma1}.
Using the expressions from the ansatz~\eqref{gZ1}--\eqref{gbasemetric}, together with the self-duality of $d\bar\beta^\a$ and the anti-self-duality of $d\bar\omega^\a$, we find
\bea
d\omega+*_4d\omega &=& (d\bar Z^\a_1+d\bar Z^\a_2)\wedge \dot{f}^\a_i\,dx^i + *_4[(d\bar Z^\a_1+d\bar Z^\a_2)\wedge \dot{f}^\a_i\,dx^i] + 2\,d\bar\beta^\a\,|\dot{f}^\a|^2\nonumber\\
&& - \partial_v [d (\bar \zeta^\a_{ki}\,\dot{f}^\a_k\,dx^i) + *_4 d (\bar \zeta^\a_{ki}\,\dot{f}^\a_k\,dx^i)]\,.
\eea
The terms in the second line can be simplified with the help of the identity (\ref{identityzeta}) and the fact that $\bar \zeta$ is anti-self-dual, obtaining
\bea
d\omega+*_4d\omega &=& (d\bar Z^\a_1+d\bar Z^\a_2)\wedge \dot{f}^\a_i\,dx^i + *_4[(d\bar Z^\a_1+d\bar Z^\a_2)\wedge \dot{f}^\a_i\,dx^i] + 2\,d\bar\beta^\a\,|\dot{f}^\a|^2\nonumber\\
&&- \partial_v [\bar\beta^\a\wedge \dot{f}^\a_i\,dx^i + *_4 d (\bar\beta^\a\wedge \dot{f}^\a_i\,dx^i)]\,.
\eea
Subtracting from $d\omega+*_4 d\omega$ the expression~\eqref{eq:theta1minuspsi} for $\Theta_1-\psi$, we find, according to~\eqref{eq:omegaeqlin} 
\be
\Theta_2-\psi =  d({\bar Z}^\a_1 \,\dot{f}^\a_i\,dx^i ) + *_4 d({\bar Z}^\a_1 \,\dot{f}^\a_i\,dx^i) +d\bar\beta^\a |\dot{f}^\a|^2 -\frac{1}{2}\partial_v [\bar\beta^\a\wedge \dot{f}^\a_i\,dx^i + *_4 (\bar\beta^\a\wedge \dot{f}^\a_i\,dx^i)] \,,
\ee
which is of the same form as $\Theta_1-\psi$ with the exchange of $Z_1$ with $Z_2$; it immediately follows that the 1-form $a_2$ exists and it is given by
\be
a_2=({\bar Z}^\a_1-1)\,\dot{f}^\a_i\,dx^i - {\bar \gamma}^\a_{1\,ik}\,\dot{f}^\a_k\,dx^i + \bar \beta^\a\,|\dot{f}^\a|^2\,.
\ee

The last equation to be verified is the one for $\mathcal{F}$, given in~\eqref{eq:Rvvlin}. From the ansatz~~\eqref{gZ1}--\eqref{gbasemetric} we see that $\mathcal{F}$ is a linear combination of
$\bar Z^\a_1$, $\bar Z^\a_2$ and $\bar\omega^\a$, which are harmonic according to~\eqref{eq:D1D5ha}: hence $d *_4 d\mathcal{F}=0$, and on the l.h.s. of equation~\eqref{eq:Rvvlin} only the term containing $\dot{\omega}$ contributes. Thus the l.h.s. of~\eqref{eq:Rvvlin} is
\bea
*_4 d *_4 \dot{\omega} =*_4\partial_v [d *_4 ((\bar Z^\a_1+\bar Z^\a_2)\,\dot{f}^\a_i\,dx^i) -\partial_v (d *_4 (\bar \zeta^\a_{ki}\,\dot{f}^\a_k\,dx^i))]\,,
\eea
where a term proportional to $d *_4 \bar\omega^\a$ and one proportional to $d *_4 \bar\beta^\a$
have been dropped on account of~\eqref{eq:D1D5ha}. One has
\be
*_4 d *_4 ((\bar Z^\a_1+\bar Z^\a_2)\,\dot{f}^\a_i\,dx^i) = -\partial_i (\bar Z^\a_1+\bar Z^\a_2) \,\dot{f}^\a_i = \partial_v (\bar Z^\a_1+\bar Z^\a_2)\,,
\ee
and
\be
*_4d *_4 (\bar \zeta^\a_{ki}\,\dot{f}^\a_k\,dx^i) = -\partial_k\, \bar\zeta^\a_{ik}\,\dot{f}^\a_i = -\bar\beta^\a_i\,\dot{f}^\a_i\,,
\ee
where we have used that $\bar\beta^\a = -*_4 d\bar\zeta^\a= *_4\,d *_4 \bar\zeta^\a$, as it follows from~\eqref{eq:D1D5ha}; hence
\be
*_4 d *_4 \dot{\omega}  =\partial_v^2 (\bar Z^\a_1+\bar Z^\a_2 +\bar\beta^\a_i\,\dot{f}^\a_i)\,.
\ee
One the r.h.s. of~\eqref{eq:Rvvlin} one finds
\be
 \partial_v^2 (Z_1 + Z_2)+\frac{1}{2}\,\partial_v^2 (h_{ii}) = \partial_v^2 (\bar Z^\a_1+\bar Z^\a_2 + 2\,\bar\beta^\a_i\,\dot{f}^\a_i - \bar\beta^\a_i\,\dot{f}^\a_i) =*_4 d *_4 \dot{\omega} \,,
\ee
which proves~\eqref{eq:Rvvlin}.

\section{Coordinate shift}
\label{Appb}
Let us consider the supergravity ansatz of Section~\ref{sec:ansatz}
with a flat 4D base metric $h_{ij} =\delta_{ij}$. We perform the shift
$x^i \to x^i - f^i(v)$ on the $R^4$ coordinates and rewrite the
resulting 10D metric in the form dictated by the
ansatz~\eqref{metrica}. As a result we obtain a new form for the 4D
base metric and the other geometric data
\bea
ds^2_4 &=& (1-\bar\beta'_k\,\dot{f}_k)\,dx^i dx^i + (\bar\beta'_j\,\dot{f}_i +\bar\beta'_i\,\dot{f}_j )\,dx^i dx^j + \frac{\bar\beta'_i \,\bar\beta'_j}{1-\bar\beta'_k\,\dot{f}_k}|\dot{f}|^2 \,dx^i dx^j,\\
\beta &=& \frac{\bar\beta'}{1-\bar\beta'_k \, \dot{f}_k}\,,\\
Z_I&=&\frac{\bar{Z}'_I}{1-\bar\beta'_k\,\dot{f}_k}\quad I=1,2,4\,,\\
\omega&=& \bar\omega' + \bar\beta'\, \Bigl(\frac{{\bar\omega}'_l \,\dot{f}_l}{1-\bar\beta'_k\, \dot{f}_k}+\frac{\bar{Z}'_1 {Z}'_2}{\bar\alpha'\,(1-\bar\beta'_k\, \dot{f}_k)^2}\,|\dot{f}|^2\Bigr)+ \frac{\bar{Z}'_1 \bar{Z}'_2}{\bar\alpha'\,(1-\bar\beta_k\, \dot{f}_k)}\,\dot{f}_i\,dx^i\,,\\
\mathcal{F} &=& \bar{\mathcal{F}'}\,(1-\bar\beta'_k\, \dot{f}_k) - 2\,\bar{\omega}'_k \, \dot{f}_k -\frac{\bar{Z}'_1 \bar{Z}'_2}{\bar\alpha'\,(1-\bar\beta'_k\,\dot{f}'_k)}\,|\dot{f}^2|\,,
\eea 
where the quantities on the l.h.s. define the solution after the
shift, while the barred and primed quantities on the r.h.s. are the
original geometric data (i.e. those in the frame where the base metric
is flat) evaluated at the point $x-f^i(v)$. By repeating the same change of
variables on the other supergravity fields we obtain
\bea
 a_1 &=& (1-\bar\beta'_k \,\dot{f}_k)\,a'_1 + \bar\beta'\,\bar{a}'_{1 k}\,\dot{f}_k + \bar{Z}_2\Bigl(\dot{f}_i \,dx^i + \frac{\bar\beta'}{1-\bar\beta'_k\,\dot{f}_k}\,|\dot{f}^2|\Bigr) - \bar\gamma'_{2 ij}\,dx^i \,\dot{f}^j\,,\\
a_4 &=& (1-\bar\beta'_k \,\dot{f}_k)\,\bar{a}_4 + \bar\beta'\,\bar{a}'_{4 k}\,\dot{f}_k + \bar{Z}_4\Bigl(\dot{f}_i \,dx^i + \frac{\bar\beta'}{1-\bar\beta'_k\,\dot{f}_k}\,|\dot{f}|^2\Bigr) - \bar\delta'_{2 ij}\,dx^i \,\dot{f}^j\,,\\
\gamma_2&=&\bar\gamma'_2 + \bar\gamma'_{2 ij}\,\dot{f}_i\,\frac{\bar\beta'}{1-\bar\beta'_k\,\dot{f}_k}\wedge dx^j\,,\\
\delta_2&=&\bar\delta'_2 + \bar\delta'_{2 ij}\,\dot{f}_i\,\frac{\bar\beta'}{1-\bar\beta'_k\,\dot{f}_k}\wedge dx^j\,,\\
x_3&=& \bar{x}'_3\,(1-\bar\beta'_k\,\dot{f}_k)+\frac{1}{2}\, \bar\beta'\wedge
\bar{x}'_{3\,ijk}\,\dot{f}_k\,dx^i\wedge dx^j+\bar{Z}'_4\,\bar\gamma'_2\wedge
\Bigl(\dot{f}_i\,dx^i +
\frac{\bar\beta'}{1-\bar\beta'_k\,\dot{f}_k}\,|\dot{f}|^2\Bigr)\nonumber\\
&&+ \bar{Z}'_4\, \bar\gamma'_{2\,ij}\,\dot{f}_i\,dx^j\wedge \frac{\bar\beta'}{1-\bar\beta'_k\,\dot{f}_k} \wedge \dot{f}_k\,dx^k\,.
\eea
It is straightforward to check that
Eqs.~\eqref{D1D5exactfirst}--\eqref{D1D5exactbase} are reproduced by
choosing the quantities on the r.h.s. as done
in~\eqref{flatZ1}--\eqref{flatbasemetric} and then by linearizing the
result in $\beta^{\scriptscriptstyle D1D5}$ and $\omega^{\scriptscriptstyle D1D5}$. As discussed in
Section~\ref{sec:BGSW}, this keeps the dependence on the D1 and D5
charges $Q_1$ and $Q_5$ exact and includes only the first order
backreaction due to the KK-monopole dipole charge $\beta^{\scriptscriptstyle D1D5}$, to $\omega^{\scriptscriptstyle D1D5}$
and to the other objects that are related to $\beta^{\scriptscriptstyle D1D5}$ or $\omega^{\scriptscriptstyle D1D5}$ by
the supergravity equations.

\bibliographystyle{utphys}      

\begin{thebibliography}{10}

\bibitem{Strominger:1996sh}
A.~Strominger and C.~Vafa, ``{Microscopic origin of the Bekenstein-Hawking
  entropy},'' \href{http://dx.doi.org/10.1016/0370-2693(96)00345-0}{{\em
  Phys.Lett.} {\bfseries B379} (1996) 99--104},
  \href{http://arxiv.org/abs/hep-th/9601029}{{\ttfamily arXiv:hep-th/9601029
  [hep-th]}}.

\bibitem{Callan:1996dv}
C.~G. Callan and J.~M. Maldacena, ``{D-brane approach to black hole quantum
  mechanics},'' \href{http://dx.doi.org/10.1016/0550-3213(96)00225-8}{{\em
  Nucl.Phys.} {\bfseries B472} (1996) 591--610},
  \href{http://arxiv.org/abs/hep-th/9602043}{{\ttfamily arXiv:hep-th/9602043
  [hep-th]}}.

\bibitem{Lunin:2001fv}
O.~Lunin and S.~D. Mathur, ``{Metric of the multiply wound rotating string},''
  \href{http://dx.doi.org/10.1016/S0550-3213(01)00321-2}{{\em Nucl. Phys.}
  {\bfseries B610} (2001) 49--76},
\href{http://arxiv.org/abs/hep-th/0105136}{{\ttfamily arXiv:hep-th/0105136}}.

\bibitem{Lunin:2001jy}
O.~Lunin and S.~D. Mathur, ``{AdS/CFT duality and the black hole information
  paradox},'' \href{http://dx.doi.org/10.1016/S0550-3213(01)00620-4}{{\em Nucl.
  Phys.} {\bfseries B623} (2002) 342--394},
\href{http://arxiv.org/abs/hep-th/0109154}{{\ttfamily arXiv:hep-th/0109154}}.

\bibitem{Lunin:2002qf}
O.~Lunin and S.~D. Mathur, ``{Statistical interpretation of Bekenstein entropy
  for systems with a stretched horizon},''
  \href{http://dx.doi.org/10.1103/PhysRevLett.88.211303}{{\em Phys. Rev. Lett.}
  {\bfseries 88} (2002) 211303},
\href{http://arxiv.org/abs/hep-th/0202072}{{\ttfamily arXiv:hep-th/0202072}}.

\bibitem{Mathur:2005zp}
S.~D. Mathur, ``{The fuzzball proposal for black holes: An elementary
  review},'' \href{http://dx.doi.org/10.1002/prop.200410203}{{\em Fortsch.
  Phys.} {\bfseries 53} (2005) 793--827},
\href{http://arxiv.org/abs/hep-th/0502050}{{\ttfamily arXiv:hep-th/0502050}}.

\bibitem{Bena:2007kg}
I.~Bena and N.~P. Warner, ``{Black holes, black rings and their microstates},''
  \href{http://dx.doi.org/10.1007/978-3-540-79523-0}{{\em Lect. Notes Phys.}
  {\bfseries 755} (2008) 1--92},
\href{http://arxiv.org/abs/hep-th/0701216}{{\ttfamily arXiv:hep-th/0701216}}.

\bibitem{Skenderis:2008qn}
K.~Skenderis and M.~Taylor, ``{The fuzzball proposal for black holes},''
  \href{http://dx.doi.org/10.1016/j.physrep.2008.08.001}{{\em Phys. Rept.}
  {\bfseries 467} (2008) 117--171},
\href{http://arxiv.org/abs/0804.0552}{{\ttfamily arXiv:0804.0552 [hep-th]}}.

\bibitem{Mathur:2008nj}
S.~D. Mathur, ``{Fuzzballs and the information paradox: a summary and
  conjectures},''
\href{http://arxiv.org/abs/0810.4525}{{\ttfamily arXiv:0810.4525 [hep-th]}}.

\bibitem{Balasubramanian:2008da}
V.~Balasubramanian, J.~de~Boer, S.~El-Showk, and I.~Messamah, ``{Black Holes as
  Effective Geometries},''
  \href{http://dx.doi.org/10.1088/0264-9381/25/21/214004}{{\em
  Class.Quant.Grav.} {\bfseries 25} (2008) 214004},
  \href{http://arxiv.org/abs/0811.0263}{{\ttfamily arXiv:0811.0263 [hep-th]}}.

\bibitem{Chowdhury:2010ct}
B.~D. Chowdhury and A.~Virmani, ``{Modave Lectures on Fuzzballs and Emission
  from the D1-D5 System},''
\href{http://arxiv.org/abs/1001.1444}{{\ttfamily arXiv:1001.1444 [Unknown]}}.

\bibitem{Mathur:2012zp}
S.~D. Mathur, ``{Black Holes and Beyond},''
  \href{http://dx.doi.org/10.1016/j.aop.2012.05.001}{{\em Annals Phys.}
  {\bfseries 327} (2012) 2760--2793},
\href{http://arxiv.org/abs/1205.0776}{{\ttfamily arXiv:1205.0776 [hep-th]}}.

\bibitem{Lunin:2004uu}
O.~Lunin, ``{Adding momentum to D1-D5 system},''
  \href{http://dx.doi.org/10.1088/1126-6708/2004/04/054}{{\em JHEP} {\bfseries
  04} (2004) 054},
\href{http://arxiv.org/abs/hep-th/0404006}{{\ttfamily arXiv:hep-th/0404006}}.

\bibitem{Giusto:2004id}
S.~Giusto, S.~D. Mathur, and A.~Saxena, ``{Dual geometries for a set of
  3-charge microstates},''
  \href{http://dx.doi.org/10.1016/j.nuclphysb.2004.09.001}{{\em Nucl. Phys.}
  {\bfseries B701} (2004) 357--379},
\href{http://arxiv.org/abs/hep-th/0405017}{{\ttfamily arXiv:hep-th/0405017}}.

\bibitem{Giusto:2004ip}
S.~Giusto, S.~D. Mathur, and A.~Saxena, ``{3-charge geometries and their CFT
  duals},'' \href{http://dx.doi.org/10.1016/j.nuclphysb.2005.01.009}{{\em Nucl.
  Phys.} {\bfseries B710} (2005) 425--463},
\href{http://arxiv.org/abs/hep-th/0406103}{{\ttfamily arXiv:hep-th/0406103}}.

\bibitem{Bena:2005va}
I.~Bena and N.~P. Warner, ``{Bubbling supertubes and foaming black holes},''
  \href{http://dx.doi.org/10.1103/PhysRevD.74.066001}{{\em Phys.Rev.}
  {\bfseries D74} (2006) 066001},
  \href{http://arxiv.org/abs/hep-th/0505166}{{\ttfamily arXiv:hep-th/0505166
  [hep-th]}}.

\bibitem{Berglund:2005vb}
P.~Berglund, E.~G. Gimon, and T.~S. Levi, ``{Supergravity microstates for BPS
  black holes and black rings},''
  \href{http://dx.doi.org/10.1088/1126-6708/2006/06/007}{{\em JHEP} {\bfseries
  0606} (2006) 007}, \href{http://arxiv.org/abs/hep-th/0505167}{{\ttfamily
  arXiv:hep-th/0505167 [hep-th]}}.

\bibitem{Ford:2006yb}
J.~Ford, S.~Giusto, and A.~Saxena, ``{A class of BPS time-dependent 3-charge
  microstates from spectral flow},''
  \href{http://dx.doi.org/10.1016/j.nuclphysb.2007.09.008}{{\em Nucl. Phys.}
  {\bfseries B790} (2008) 258--280},
\href{http://arxiv.org/abs/hep-th/0612227}{{\ttfamily arXiv:hep-th/0612227}}.

\bibitem{Giusto:2009qq}
S.~Giusto, J.~F. Morales, and R.~Russo, ``{D1D5 microstate geometries from
  string amplitudes},'' \href{http://dx.doi.org/10.1007/JHEP03(2010)130}{{\em
  JHEP} {\bfseries 03} (2010) 130},
\href{http://arxiv.org/abs/0912.2270}{{\ttfamily arXiv:0912.2270 [hep-th]}}.

\bibitem{Black:2010uq}
W.~Black, R.~Russo, and D.~Turton, ``{The supergravity fields for a D-brane
  with a travelling wave from string amplitudes},''
  \href{http://dx.doi.org/10.1016/j.physletb.2010.09.059}{{\em Phys. Lett.}
  {\bfseries B694} (2010) 246--251},
\href{http://arxiv.org/abs/1007.2856}{{\ttfamily arXiv:1007.2856 [hep-th]}}.

\bibitem{Giusto:2011fy}
S.~Giusto, R.~Russo, and D.~Turton, ``{New D1-D5-P geometries from string
  amplitudes},'' \href{http://dx.doi.org/10.1007/JHEP11(2011)062}{{\em JHEP}
  {\bfseries 1111} (2011) 062},
\href{http://arxiv.org/abs/1108.6331}{{\ttfamily arXiv:1108.6331 [hep-th]}}.

\bibitem{Giusto:2012gt}
S.~Giusto and R.~Russo, ``{Adding new hair to the 3-charge black ring},''
  \href{http://dx.doi.org/10.1088/0264-9381/29/8/085006}{{\em
  Class.Quant.Grav.} {\bfseries 29} (2012) 085006},
\href{http://arxiv.org/abs/1201.2585}{{\ttfamily arXiv:1201.2585 [hep-th]}}.

\bibitem{deBoer:2010ud}
J.~de~Boer and M.~Shigemori, ``{Exotic branes and non-geometric backgrounds},''
  \href{http://dx.doi.org/10.1103/PhysRevLett.104.251603}{{\em Phys.Rev.Lett.}
  {\bfseries 104} (2010) 251603},
\href{http://arxiv.org/abs/1004.2521}{{\ttfamily arXiv:1004.2521 [hep-th]}}.

\bibitem{deBoer:2012ma}
J.~de~Boer and M.~Shigemori, ``{Exotic Branes in String Theory},''
\href{http://arxiv.org/abs/1209.6056}{{\ttfamily arXiv:1209.6056 [hep-th]}}.

\bibitem{Bena:2011uw}
I.~Bena, J.~de~Boer, M.~Shigemori, and N.~P. Warner, ``{Double, Double
  Supertube Bubble},'' \href{http://dx.doi.org/10.1007/JHEP10(2011)116}{{\em
  JHEP} {\bfseries 1110} (2011) 116},
\href{http://arxiv.org/abs/1107.2650}{{\ttfamily arXiv:1107.2650 [hep-th]}}.

\bibitem{Bena:2011dd}
I.~Bena, S.~Giusto, M.~Shigemori, and N.~P. Warner, ``{Supersymmetric Solutions
  in Six Dimensions: A Linear Structure},''
  \href{http://dx.doi.org/10.1007/JHEP03(2012)084}{{\em JHEP} {\bfseries 1203}
  (2012) 084},
\href{http://arxiv.org/abs/1110.2781}{{\ttfamily arXiv:1110.2781 [hep-th]}}.

\bibitem{Niehoff:2012wu}
B.~E. Niehoff, O.~Vasilakis, and N.~P. Warner, ``{Multi-Superthreads and
  Supersheets},''
\href{http://arxiv.org/abs/1203.1348}{{\ttfamily arXiv:1203.1348 [hep-th]}}.

\bibitem{Gutowski:2003rg}
J.~B. Gutowski, D.~Martelli, and H.~S. Reall, ``{All Supersymmetric solutions
  of minimal supergravity in six- dimensions},''
  \href{http://dx.doi.org/10.1088/0264-9381/20/23/008}{{\em Class.Quant.Grav.}
  {\bfseries 20} (2003) 5049--5078},
\href{http://arxiv.org/abs/hep-th/0306235}{{\ttfamily arXiv:hep-th/0306235
  [hep-th]}}.

\bibitem{Cariglia:2004kk}
M.~Cariglia and O.~A. Mac~Conamhna, ``{The General form of supersymmetric
  solutions of N=(1,0) U(1) and SU(2) gauged supergravities in
  six-dimensions},'' \href{http://dx.doi.org/10.1088/0264-9381/21/13/006}{{\em
  Class.Quant.Grav.} {\bfseries 21} (2004) 3171--3196},
\href{http://arxiv.org/abs/hep-th/0402055}{{\ttfamily arXiv:hep-th/0402055
  [hep-th]}}.

\bibitem{GMPR}
S.~Giusto, L.~Martucci, M.~Petrini, and R.~Russo, in preparation.

\bibitem{DiVecchia:1999rh}
P.~Di~Vecchia and A.~Liccardo, ``{D branes in string theory. I},'' {\em NATO
  Adv. Study Inst. Ser. C. Math. Phys. Sci.} {\bfseries 556} (2000) 1--59,
\href{http://arxiv.org/abs/hep-th/9912161}{{\ttfamily arXiv:hep-th/9912161}}.

\bibitem{DiVecchia:1999fx}
P.~Di~Vecchia and A.~Liccardo, ``{D-branes in string theory. 2.},''
\href{http://arxiv.org/abs/hep-th/9912275}{{\ttfamily arXiv:hep-th/9912275
  [hep-th]}}.

\bibitem{Hikida:2003bq}
Y.~Hikida, H.~Takayanagi, and T.~Takayanagi, ``{Boundary states for D-branes
  with traveling waves},'' {\em JHEP} {\bfseries 04} (2003) 032,
\href{http://arxiv.org/abs/hep-th/0303214}{{\ttfamily arXiv:hep-th/0303214}}.

\bibitem{Blum:2003if}
J.~D. Blum, ``{Gravitational radiation from travelling waves on D- strings},''
  \href{http://dx.doi.org/10.1103/PhysRevD.68.086003}{{\em Phys. Rev.}
  {\bfseries D68} (2003) 086003},
\href{http://arxiv.org/abs/hep-th/0304173}{{\ttfamily arXiv:hep-th/0304173}}.

\bibitem{Bachas:2003sj}
C.~P. Bachas and M.~R. Gaberdiel, ``{World-sheet duality for D-branes with
  travelling waves},'' {\em JHEP} {\bfseries 03} (2004) 015,
\href{http://arxiv.org/abs/hep-th/0310017}{{\ttfamily arXiv:hep-th/0310017}}.

\bibitem{Callan:1995hn}
C.~G. Callan, J.~M. Maldacena, and A.~W. Peet, ``{Extremal Black Holes As
  Fundamental Strings},''
  \href{http://dx.doi.org/10.1016/0550-3213(96)00315-X}{{\em Nucl. Phys.}
  {\bfseries B475} (1996) 645--678},
\href{http://arxiv.org/abs/hep-th/9510134}{{\ttfamily arXiv:hep-th/9510134}}.

\bibitem{Dabholkar:1995nc}
A.~Dabholkar, J.~P. Gauntlett, J.~A. Harvey, and D.~Waldram, ``{Strings as
  Solitons \& Black Holes as Strings},''
  \href{http://dx.doi.org/10.1016/0550-3213(96)00266-0}{{\em Nucl. Phys.}
  {\bfseries B474} (1996) 85--121},
\href{http://arxiv.org/abs/hep-th/9511053}{{\ttfamily arXiv:hep-th/9511053}}.

\bibitem{Berkooz:1996km}
M.~Berkooz, M.~R. Douglas, and R.~G. Leigh, ``{Branes intersecting at
  angles},'' \href{http://dx.doi.org/10.1016/S0550-3213(96)00452-X}{{\em
  Nucl.Phys.} {\bfseries B480} (1996) 265--278},
\href{http://arxiv.org/abs/hep-th/9606139}{{\ttfamily arXiv:hep-th/9606139
  [hep-th]}}.

\bibitem{Bachas:1995kx}
C.~Bachas, ``{D-brane dynamics},''
  \href{http://dx.doi.org/10.1016/0370-2693(96)00238-9}{{\em Phys.Lett.}
  {\bfseries B374} (1996) 37--42},
\href{http://arxiv.org/abs/hep-th/9511043}{{\ttfamily arXiv:hep-th/9511043
  [hep-th]}}.

\bibitem{Kanitscheider:2007wq}
I.~Kanitscheider, K.~Skenderis, and M.~Taylor, ``{Fuzzballs with internal
  excitations},'' {\em JHEP} {\bfseries 06} (2007) 056,
\href{http://arxiv.org/abs/0704.0690}{{\ttfamily arXiv:0704.0690 [hep-th]}}.

\bibitem{Garfinkle:1990jq}
D.~Garfinkle and T.~Vachaspati, ``{Cosmic string traveling waves},''
\href{http://dx.doi.org/10.1103/PhysRevD.42.1960}{{\em Phys.Rev.} {\bfseries
  D42} (1990) 1960--1963}.

\end{thebibliography}

\providecommand{\href}[2]{#2}\begingroup\raggedright\endgroup

\end{document}